\documentclass{article}

\usepackage{PRIMEarxiv}

\usepackage[utf8]{inputenc} 
\usepackage[T1]{fontenc}    
\usepackage{hyperref}       
\usepackage{url}            
\usepackage{booktabs}       
\usepackage{amsfonts}       
\usepackage{nicefrac}       
\usepackage{microtype}      
\usepackage{lipsum}
\usepackage{graphicx}
\usepackage{natbib}
\usepackage{amsmath}
\usepackage{float}
\graphicspath{{media/}}     

\usepackage{xcolor}
\definecolor{magenta}{RGB}{255, 50, 255}
\definecolor{darkgoldenrod2}{RGB}{245, 174, 14}
\definecolor{purple-blue}{RGB}{15, 10, 120}
\definecolor{purple}{RGB}{160, 30, 100}
\definecolor{skyblue}{RGB}{135, 216, 235}
\definecolor{tealgreen}{RGB}{126, 236, 200}
\definecolor{leafgreen}{RGB}{79, 199, 0}

\title{wdiexplorer: An R package Designed for Exploratory Analysis of World Development Indicators (WDI) Data
\thanks{\textit{\underline{Citation}}: 
\textbf{Authors. Title. Pages.... DOI:000000/11111.}} 
}

\author{
  Oluwayomi Akinfenwa\\
  Hamilton Institute\\
  Maynooth University\\
  Maynooth\\
  Co. Kildare, Ireland \\ 
  \texttt{oluwayomiakinfenwa@gmail.com} \\
   \And
   Niamh Cahill\\
  Department of Mathematics and Statistics\\
  Maynooth University\\
  Maynooth\\
  Co. Kildare, Ireland \\
  \texttt{Niamh.Cahill@mu.ie} \\
  \And
 Catherine Hurley\\
  Department of Mathematics and Statistics\\
  Maynooth University\\
  Maynooth\\
  Co. Kildare, Ireland \\
  \texttt{Catherine.Hurley@mu.ie} \\
}

\begin{document}
\maketitle

\begin{abstract}
The World Development Indicators (WDI) database provides a wide range of global development data, maintained and published by the World Bank. Our \textit{wdiexplorer} package offers a comprehensive workflow that sources WDI data via the \textit{WDI} R package, prepares and explores country-level panel data of the WDI through computational functions to calculate diagnostic metrics and visualise the outputs. By leveraging the functionalities of \textit{wdiexplorer} package, users can efficiently explore any indicator dataset of the WDI, compute diagnostic indices, and visualise the metrics by incorporating the pre-defined grouping structures to identify patterns, outliers, and other interesting features of temporal behaviours. This paper presents the \textit{wdiexplorer} package, demonstrates its functionalities using the WDI: PM$_{2.5}$ air pollution dataset, and discusses the observed patterns and outliers across countries and within groups of country-level panel data.
\end{abstract}


\section{Introduction}
The World Development Indicators (WDI, ``the world bank collection of development indicators'') \citep{worldbank-wdi}, is a database maintained and published by the World Bank that provides a wide range of global development data. It includes economic, social, demographic and environmental indicators, sourced from 60 international databases. The five largest sources are Education Statistics, the Atlas of Social Protection: Indicators of Resilience and Equity, Quarterly External Debt Statistics, the Disability Data Hub, and the World Development Indicators, with Education Statistics contributing the highest number of indicators (7,182). These data are standardised by the World Bank to ensure consistency across countries and over time. While most WDI series began in $1960$, coverage varies by indicator and country, with many indicators, particularly those related to the sustainable development goals, and renewable energy adoption, starting in the $1990$s and later years. Data are predominantly collected on an annual basis. However, the reporting frequency varies by indicator, with some recorded biennially or triennially, and some at quarterly intervals.

The WDI data are structured as country-level panel data, consisting of repeated observations of the same variables collected over time from multiple countries. This structured and consistent format captures both changes within individual countries and variation across countries over time. However, the underlying temporal patterns across countries are not easily identified. To facilitate the exploration of WDI country-level panel data, the WDI database website provides an online interactive tool that allows users to visualise the temporal behaviour of data series across countries. Despite this functionality, overlapping visuals often create complexity highlighting the need for enhanced exploratory visualisation tools. Such tools should allow users to explore data in ways that uncover hidden patterns, identify outliers and other interesting characteristics of the data, and also enhance better interpretation of temporal behaviours.

This paper introduces the \textit{wdiexplorer} R package, a set of tools designed to enhance the exploratory analysis of country-level panel data of the WDI. In addition to the country-level information of the WDI dataset, it includes pre-defined grouping variables such as region, income and lending category. The \textit{wdiexplorer} explicitly incorporate these groupings into the exploration of country-level panel data. It computes group-based measures that characterise panel data behaviour and produces interactive visualisations to identify patterns, temporal dynamics and shape-based features within and across countries. This package also helps to identify outliers and structural variations that might otherwise remain hidden when examining countries individually or as a global aggregate. It facilitates these by enabling comparisons of each country with others in the same pre-defined group, \textit{wdiexplorer} supports a more context-aware exploration of development trends. The package is currently available on GitHub at \url{https://github.com/Oluwayomi-Olaitan/wdiexplorer/}.

The remaining sections of this paper are organised as follows. Section 2 provides an overview of the background to this study, outlining the importance of incorporating grouping structures in country-level panel data exploration, reviewing existing exploratory tools, and examining summaries of time series data. Section 3 details our methodology, which extends scagnostics concepts and some established time series measures to exploratory analysis workflow through a collection of diagnostic indices that characterise the key behaviours of the data series. In Section 4, we describe the functions of the 
\textit{wdiexplorer} package. The utility of the package is illustrated using the mean annual exposure levels to ambient PM$_{2.5}$ air pollution indicator data of the WDI in Section 5. Finally, in Section 6, we conclude by evaluating the benefits of \textit{wdiexplorer} package, as well as other related future work.

\section{Background}\label{background}

In this section, we discuss WDI data. We highlight grouping variables present in the data set, alongside the main indicator of interest, that can enhance meaningful interpretation and comparison of country-level behavioural patterns. We also review existing exploratory tools for analysing country-level data, with a focus on identifying patterns, outliers, and other characteristics within the data set.

\subsection{World Development Indicators (WDI) Data}

In R, World Bank-hosted databases are accessible through several packages, including \textit{wbstats} \citep{wbstats}, \textit{WDI} \citep{WDI}, and \textit{worldbank} \citep{worldbank}, each offering slightly different approaches to sourcing and downloading the World Bank data. This paper focuses on the \textit{WDI} package to source and download data, which provides access to 22,806 indicators across 296 reporting entities. These entities include not only countries, but also territories, regional aggregates, and income-based aggregates. According to \cite{worldbank_classifier}, there are currently 189 World Bank member countries along with 28 other territories with populations over 30,000. After excluding non-country aggregates from the full list of WDI reporting entities, 217 distinct country and territory entities remain with individual records.

The WDI data contain additional variables that capture geographic and economic factors of each country and territory entities. They are classified by the World Bank into 7 geographic regions: East Asia \& Pacific, Europe \& Central Asia, Latin America \& Caribbean, Middle East \& North Africa, North America, South Asia, and Sub-Saharan Africa. The World Bank also divides their economies into four income groups: low income, lower-middle income, upper-middle income, and high income. This classification is updated annually on the 1st of July, based on the Gross National Income (GNI) per capita from the previous calendar year. Once assigned, a country’s income group remains fixed for the entire fiscal year until the next update. Countries are classified into three lending categories: International Development Association (IDA), International Bank for Reconstruction and Development (IBRD), and Blend based on the operational policies of the World Bank. IDA countries are low income and qualify for financial loans with low to no interest. IBRD countries are middle-income and have access to standard market-based loans while Blend countries are eligible for both IDA loans and standard IBRD loans \citep{worldbank_country_lending}.

Given the country-level panel structure of the WDI data, the pre-defined grouping information available in the data set (region, income, and lending) groupings provides essential information that can be utilised to enhance more meaningful comparisons of country behavioural patterns. \cite{gelman2007data} highlight that grouping structures in country-level panel data are widely recognised in both exploratory and modelling analyses, and emphasise that overlooking grouping structures, such as countries nested within geographic and socioeconomic groupings can lead to oversimplified or inaccurate interpretation of the data. Explicitly accounting for inherent groupings facilitates the identification of meaningful patterns that might be hidden when such structures are ignored. It also provdies a clearer understanding of group-wise patterns, highlights sources of variation, and supports the detection of outliers. For example, interpreting Ireland's birth rate over time is more informative when explored in relation to other European countries, where shared policy may shape similar demographic trends, rather than comparing it to countries in geographically and socio-economically distinct regions, like Sub-Saharan Africa or South Asia. 

Grouping structures offer two crucial comparative perspectives: understanding countries relative to other countries in their geographical or socioeconomic groups, and identifying group-wise patterns across levels, facilitating comparison between groups. Guided by these perspectives, our approach to explore country-level panel data is organised around two complementary strategies: (1) a collection of diagnostic indices that characterise panel data behaviour, (2) group-informed exploration of country-level panel data that leverage the pre-defined groupings of the data through interactive visuals to capture behavioural patterns and highlight group-based features.

\subsection{Country-level Panel Data Exploration}

Exploration of country-level panel data is the process of revealing patterns, identifying outliers and other structural behaviours among observational units across countries and over time. Traditional approach of exploring country-level panel data often rely on line plots of series trajectories organised by grouping variables. However, these can become overwhelming and insufficient to capture complex temporal and group structures in large data sets. While ensemble graphics provide an effective way to explore multiple faceted temporal data, \cite{RJ-2021-050} discuss the challenges of managing such complexity during exploratory analysis, particularly the difficulty in visualising grouping structures and temporal dynamics simultaneously. To address this challenge, \cite{RJ-2021-050} developed the \textit{tsibbletalk} \citep{tsibbletalk} R package, which leverage the \textit{tsibble}  data structure \citep{tsibble} that produces a tidy format of time series and panel data as index and key for series identification. The \textit{tsibbletalk} package is integrated with the \textit{crosstalk}  R package \citep{crosstalk} that enables filtering of data views and highlighting in html-widgets without relying on a Shiny app, to provide linked brushing by hovering across multiple coordinated views to make grouping structures interactive and visually explorable. 

Several other \textit{R} packages also provide valuable tools for exploring time series and panel data. For example, 
\textit{brolgar}, which also uses the \textit{tsibble} data structure to produce a tidy data series, enhances the exploration of longitudinal data by providing tools to visualise individual trajectories in large datasets \citep{brolgar}. It supports sampling, stratification, and feature-based summaries of the data series, which facilitates the identification of unusual patterns within large data sets of individual series. Also, there are existing measures specifically designed to capture the temporal characteristics of time series data (e.g., \cite{hyndman2021forecasting}). These include summary statistics, autocorrelation measures, trend and seasonal strength measures, as well as tests for consistent behaviour over time. Such measures aim to uncover meaningful temporal behaviours and are implemented in the \textit{feasts} package. \textit{feasts} \cite{feasts} focuses on a concise set of features proven effective in time series analysis. \textit{tsibbletalk} and \textit{brolgar} leverage \textit{feasts} and \textit{fabletools} \citep{fabletools} respectively to summarise the temporal features of the data series but does not specifically address the integration of grouping structures into its exploratory workflow, limiting its ability to explore and compare patterns across and between groups with their temporal features. 

Although the WDI data do not strictly qualify as functional data, they exhibit comparable characteristics, consisting of observations from multiple countries over time.  \cite{qu2024exploratory} introduced visualisation tools for functional data (which they call ``exploratory functional data analysis'' or EFDA) such as rainbow plots, functional boxplots and functional bagplots and provided techniques for outlier detection and functional data clustering to detect dominant behaviours and anomalies across curves. Functional clustering, in particular, is used to partition data based on shared shape or magnitude features, supporting data-driven identification of subgroups. The rainbow plot incorporates data ordering feature that colours observations based on their order, which can reflect various indices such as time or other relevant characteristics to enhance the exploration process. However, while the \cite{qu2024exploratory} work supports identifying groupings and patterns based on shared data features, it does not explicitly incorporate pre-defined grouping structures into its framework. The \textit{wdiexplorer} package builds on its idea of rainbow plot to order series trajectories based on summary measures of the proposed collection of diagnostic indices.

If we consider WDI data as scatter plots (plots of observational units versus time) for individual countries, the idea of scagnostics (scatterplot diagnostics) \citep{tukey1985computer, wilkinson2008scagnostics} can be employed to highlight interesting relationships within the data. Scagnostics quantify structural features of two-dimensional scatterplots, such as outliers, skewness, and shape, which support data exploration across large scatterplot matrices by identifying distinctive patterns \citep{dang2014scagexplorer}. However, while scagnostics provide valuable insights into the distributional and geometric characteristics of scatterplots, they are designed for unstructured 2D relationships and do not account for structured observational units or grouping variables that are key attributes of country-level panel data.

The \textit{wdiexplorer} package adapts the conceptual framework of scagnostics measures (whilst we do not directly apply the original scagnostics measures), incorporate and extend their underlying principles alongside selected time series features from \cite{hyndman2021forecasting} (as implemented in the \textit{feasts} package), and additional measures tailored to panel data series. The package construct a collection of diagnostic indices that quantify variation, trend and shape, and sequential temporal structure across time series. Measures are calculated with a consideration of grouping structures to enhance the exploratory analysis of country-level panel data. Additionally, it refines the interactive visualisation approach used by \textit{tsibbletalk} and incorporates the concept of colouring individual units based on their ordering, as seen in the rainbow plot of EFDA, to enhance interactive visuals that summarise behavioural patterns and highlight group-based features of country-level panel data based on their metric values.

\hypertarget{methodology}{%
\section{Methodology}\label{methodology}}

In this section, we present our methodology, inspired by the scagnostics (scatterplot diagnostics) framework originated by \cite{tukey1985computer} and formalised by \cite{wilkinson2008scagnostics}, which provides a set of quantitative measures to detect anomalies in density, shape, association, and other features of 2D scatterplots. We extend scagnostics concepts with some time series features of the \textit{feasts} package to the exploratory analysis of country-level panel data through a collection of diagnostic indices that characterise key behaviours such as trends, variability and similarities of the data series. In addition, considering the heterogeneity typically present in country-level panel data, we explicitly incorporate grouping information (e.g., geographic or socioeconomic) present in the data into the exploratory process through static and interactive visuals.

\subsection{A Collection of Diagnostic Indices that Characterise Panel Data Behaviour}\label{diagnostic-indices}

Our collection of diagnostic indices are organised into three categories: (1) variation features, (2) trend and shape features, and (3) sequential temporal features. Together, these indices characterise country-level panel data and support the identification of trends, patterns, outliers, and other potentially interesting aspects of variation over time.

\begin{enumerate}
    \item \textbf{Variation Features}

The variation features capture the overall differences in the variable of interest within and between groups of countries. These include an overall dissimilarity measure for each country, a dissimilarity measure for each country relative to its pre-defined group, and an assessment of how well each country fits its group. These variation measures are formally defined below. 

\begin{itemize}
\item \textbf{Country dissimilarity} quantifies the overall dissimilarity for a country. For each country, it is calculated as the average of the Euclidean distances between that country and every other country.

    A high dissimilarity value indicates that the series for a country differs substantially from other countries, signalling potential outlier or unique behaviour, while a low dissimilarity value suggests the series closely resembles the behaviour of other countries in the dataset.

\item \textbf{Group-wise dissimilarity} measures how dissimilar each country is from others in the same group. For each country, it is calculated as the average of the Euclidean distances between that country and every other country in its pre-defined group.

\item \textbf{Silhouette width} assess how well the series for a country fits its group relative to other groups \citep{rousseeuw1987silhouettes}. For each country, it is calculated by comparing the average distance to countries within its group with the distance to the nearest neighbouring group. The silhouette width ranges from $\mathrm{-}1$ to $\mathrm{+}1$.

    A high silhouette width (close to $\mathrm{+}1$) indicates that the country is well aligned within its pre-defined group with clear separation from other groups. A silhouette width close to $0$ indicates a country lies near the boundary between two groups, and a silhouette width close to $\mathrm{-}1$ denotes a country closer to a group other that its pre-defined group.
\end{itemize}

\item \textbf{Trend and Shape Features}

In our exploratory analysis, trend and shape features describe how country-level data patterns evolve and whether they deviate from smooth, or consistent trajectories. These measures help us understand whether data changes follow steady patterns or more abrupt shifts. These are measured using trend strength, linearity, curvature and smoothness. 

Trend strength, linearity, and curvature are part of the set of time series features proposed by \cite{hyndman2021forecasting} and implemented in the \textit{feasts} package \citep{feasts}. These features are estimated using a time series decomposition \cite{cleveland1990stl}, which takes the form 

$$\; \; y_t = T_t + S_t + R_t,$$ 

for seasonal data, and 
$$\; \; y_t = T_t + R_t,$$

for non-seasonal data. $T_t$ is the smoothed trend component computed using Friedman’s SuperSmoother \citep{luedicke2015friedman}, implemented through the \texttt{stats}::\textit{supsmu} function \citep{R}, and $R_t$ is a remainder component. 

\begin{itemize}
\item \textbf{Trend strength} measures the extent to which data follows a consistent pattern over time, which could be linear or curved relative to random fluctuations \citep{hyndman2021forecasting}. This measure, which ranges from $0$ to $1$, captures how pronounced and stable the overall trend is, distinguishing steady patterns from irregular noise. As described by \cite{hyndman2021forecasting}, a data series is considered strongly trended if $\mathrm{Var}(R_t) < \mathrm{Var}(T_t +R_t)$, and \textbf{weakly trended} if $\mathrm{Var}(R_t) > \mathrm{Var}(T_t +R_t)$. Therefore, the strength of trend $F_T$ is defined as

$$F_T = \max \left(0, \; 1 - \dfrac{\mathrm{Var}(R_t)}{\mathrm{Var}(T_t + R_t)}\right).$$

\item \textbf{Linearity} measures how closely a series follows a straight-line pattern over time. It assesses whether changes occur consistently without sharp curves or bends. Following \citet{hyndman2021forecasting}, here it is defined as the coefficient of the linear term from a regression fitted to the trend component ($T$) of the decomposition, where the predictors are the first two orthogonal polynomials of the time index.

$$\mathrm{T} = \beta_0 + \beta_1 \mathrm{P}_1(t) + \beta_2 \mathrm{P}_2(t)$$
where; \\
$t$ is the time index.

$\beta_1$ is the linear term which quantifies linearity.

$\mathrm{P}_1(t)$ is the first-degree orthogonal polynomial of $t$.

$\beta_2$ is the quadratic term which quantifies curvature.

$\mathrm{P}_2(t)$ is the second-degree orthogonal polynomial of $t$.

    A high positive linearity value ($\beta_1$) indicates a strong and pronounced increasing linear trend while a high negative linearity value indicates a pronounced decreasing linear trend.

    A low positive or negative linearity value indicates a weak trend without a consistent directional pattern, while a $\beta_1$ value close to $0$ suggests the absence of any linear trend.

\item \textbf{Curvature} measures the extent to which a series deviates from a straight-line trend over time, capturing the presence of non-linear patterns; upward or downward bends. Curvature is quantified as $\beta_2$, the coefficient of the quadratic term in a polynomial regression fitted to the trend component of the decomposed series, where the predictors are the first two orthogonal polynomials of the time index \citep{hyndman2021forecasting}.

    A high positive curvature value indicates that the trend of the data series decreases and increases over time characterised by a U-shaped pattern while a high negative curvature value indicates trend increases to a peak and then declines over time characterised by an inverted U-shaped pattern.

    A low positive or negative curvature value suggests a weak U-shaped or inverted U-shaped pattern respectively, that is, the trend of the data series is not pronounced enough to be considered the dominant shape of the non-linear trend. When the curvature value is close $0$, whether slightly positive or negative, it indicates that the trend follows a nearly linear trajectory, with minimal or no curvature. 

\item \textbf{Smoothness} measures how steadily the data for each country evolves over time. Here we quantify smoothness  by the standard deviation of the lagged differences. It captures the degree of fluctuation in the trajectory of the data.

    Lower smoothness values indicate a smoother, more continuous shape, suggesting that the series evolves smoothly over time. Higher values suggest more irregular changes.

\end{itemize}

\item \textbf{Sequential Temporal Features}

In our analysis, sequential temporal features capture the order and timing of changes in the data series, focusing on patterns that unfold with time rather than the overall trend. These measures reveal persistence, oscillations, and periods of stability over time within individual series. These are: number of crossing points; longest flat spot, and autocorrelation.

Number of crossing points and longest flat spot are features that have been proven useful in time series analysis \citep{hyndman2021forecasting} and implemented in the \textit{feasts} package \cite{feasts}.  

\begin{itemize}
\item \textbf{Number of crossing points} counts how many times a data series crosses below or above its median value, capturing the extent of fluctuation around the central tendency \citep{hyndman2021forecasting}.

   A higher number of crossing points indicates frequent median crossings, signifying fluctuations above and below the median over time and reduced stability in the data series of a country. A lower number suggests more sustained behaviour in one direction, where the series tends to remain on the same side of the median for extended periods.

\item \textbf{Longest flat spot } measures the longest consecutive sequence in a series where data points lie within a single interval. This is calculated by dividing the data range into ten equal-sized intervals and identifying the maximum run length of consecutive observations that fall within the same interval \citep{hyndman2021forecasting}.

High values indicate that the series, for a given country, remains relatively unchanged, reflecting strong temporal stability, while low values suggest that the series frequently moves between different intervals.

\item \textbf{Autocorrelation} measures the correlation between each country data series and its one-period lagged values, capturing the strength and direction of association between lagged observations. The value of autocorrelation coefficient ranges from $\mathrm{-}1$ to $\mathrm{+}1$.

A high positive autocorrelation (close to $\mathrm{+}1$) indicates strong persistence, where increases (or decreases) tend to be followed by similar increases (or decreases), reflecting a smooth, consistent temporal pattern in the data series of a country. A high negative autocorrelation (close to $\mathrm{-}1$) indicates a strong inverse relationship, where increases tend to be followed by decreases (and vice versa), suggesting oscillating or alternating behaviour over time. Autocorrelation values near $0$ indicate little or no linear dependence between consecutive observations.
\end{itemize}

\end{enumerate}

The collection of diagnostic indices offers a structured summary of country-level panel data, capturing its key characteristics. While each measure provides insight on its own, their combined use with visualisation allows for a more complete view of how countries behave over time. We developed interactive visualisations to support the interpretation of individual metrics, combinations of metrics, or the full collection of the diagnostic indices.

\subsection{Static and Interactive Visuals that Summarise Behavioural Patterns and Group-Based Features} \label{interactive-visuals}

We develop a set of visual tools that connect the collection of diagnostic indices to the underlying data series in different ways. These facilitate clearer interpretation of behavioural patterns of each series, within and across groups, while helping to identify unusual or distinctive cases. 

\begin{enumerate}

 \item \textbf{The distribution plot} displays the distribution of metric values, either for all diagnostic indices or for a selected metric (one or multiple). It summarises the variation of metric values by showing the spread and shape of their distributions. There are two versions of the distribution plot: an ungrouped version, which shows the values of one or more metrics across all countries, and a grouped version, which displays the distribution within each level of a specified grouping variable. 

 \item \textbf{The partition plot} presents metric values for individual countries grouped by a specified grouping variable. The  metric value of each country is represented by a coloured bar ordered in descending order, while a lighter-shaded rectangular bar beneath indicates the average value of the metric at each group-level. This plot design is inspired by the silhouette plot \citep{rousseeuw1987silhouettes}, which quantifies and visualises how well each observation fits its assigned group. It also adapts the visual idea of presenting group averages as a shadow beneath individual values, as implemented in the \textit{flexclust} package \citep{flexclust}.

 \item \textbf{The interactive data trajectory plot} displays the trajectory of the data series for each country. It supports two modes and each mode can be rendered in two versions: ungrouped and grouped versions. Both modes display all series as uniform line plots, while the second mode highlights countries that fall within a specified percentile of any chosen diagnostic metric values. In the ungrouped version, all countries are plotted together without any grouping structure, and when highlighting is enabled, series representing countries are highlighted based on whether their metric value exceed an overall threshold calculated by the specified percentile. In the grouped version, data are faceted by a chosen grouping variable (e.g., geographic or economic groupings); when highlighting is applied, countries are highlighted based on group-specific thresholds. In both cases, the highlighted series are emphasised using a colour gradient mapped to the metric values and hovering over the lines will display the correspondence country name and metric values so that the patterns identified by the metrics can be viewed directly within the temporal behaviour of each country for a clearer understanding of what each metric captures.

 \item \textbf{The interactive parallel coordinate plot} simultaneously displays all diagnostic metrics, with each metric represented as a vertical axis. Each country is shown as an interactive line that intersects all axes, with the position along the x-axis corresponding to the diagnostic indices. To ensure comparability across metrics, all values are normalised to a scale of $0$ to $1$. Exact metric values for each country and diagnostic indices are displayed when hovering over the lines at each axis.  Similar to other plots, the interactive parallel coordinate plot supports both ungrouped and grouped versions. The ungrouped version displays all countries together as parallel lines, where lines are coloured by a grouping variable to enhance visual distinction. In the grouped version, parallel lines representing countries are faceted by a grouping variable, where countries in each group level are grouped together, and each group level are normalised to a scale of $0$ to $1$ to facilitate within-group comparison. A consistent colour scheme is applied to both views, with countries in the same group distinguished by unique colours.
 
\item \textbf{The interactive link-view of metrics and series plot} displays an interactive visualisation that connects diagnostic indices values with their corresponding series trajectories. One panel shows a scatterplot of two selected diagnostic indices (e.g., linearity and curvature), while the other shows the line plot of the data series for each country. Like the other plots, it also supports both ungrouped and grouped versions. In the ungrouped version, all countries appear in a single scatterplot and line plot panel. In the grouped version, countries are separated into facets based on a specified grouping variable (e.g., geographic or economic groupings), allowing both the scatterplot and time series panels to reflect group-specific structures. Hovering over or selecting a data point in the scatterplot highlights the corresponding trajectory with the country name, and vice versa.

\item \textbf{The missingness plot} displays a visual overview of missing data patterns across countries and years. Countries are grouped by the levels of a specified grouping variable, and the presence or absence of data is indicated by colour: missing entries are shown in black, while available data points are displayed in a light grey colour. This plot offers a structured summary of data availability, making it easy to spot countries with complete data gaps and years with widespread missingness.
\end{enumerate}

All visualisations are built using the \textit{ggplot2} \citep{ggplot2} framework, with additional support from the \textit{ggiraph} \citep{ggiraph} and \textit{ggdist} \citep{ggdist} packages. The ggiraph package enables interactivity, while the ggdist package is used to present the distributions of the diagnostic metrics.

\section{Package Structure}\label{package structure}

Our \textit{wdiexplorer} package provides a set of functions designed to facilitate the efficient exploration and visualisation of one indicator at a time of the World Development Indicators (WDI) country-level panel data. It focuses on identifying patterns, including temporal, shape-based features, outliers, and structural variations within the data, while enabling comparisons across countries and within pre-defined groups. The package functions, listed in Table \ref{functions-table}, are categorised into three groups that reflect the  stages of the exploration workflow, namely data sourcing and assessing missing data, computing diagnostic indices, and generating both static and interactive visualisations.
Examples of the usage of all of these functions are given in Section 5.

\begin{table}[H]
\centering
\caption{An overview of the wdiexplorer function names and their functionalities.}
\label{functions-table}
\begin{tabular}{lp{0.55\linewidth}} 
\toprule
Function name & Functionality \\
\midrule
\texttt{get\_wdi\_data} & Sources and locally stores data from the WDI database. \\
\texttt{plot\_missing} & Visualises missingness in the data. \\
\texttt{get\_valid\_data} & Extracts valid observations and reports gaps in country and year. \\
\midrule
\texttt{compute\_dissimilarity} & Computes Euclidean distance between pairwise countries. \\
\texttt{compute\_variation} & Computes average country dissimilarities, group-wise dissimilarities, and silhouette widths. \\
\texttt{compute\_trend\_shape\_features} & Computes trend strength, linearity, curvature, and smoothness. \\
\texttt{compute\_temporal\_features} & Computes number of crossing points, longest flat spot, and autocorrelation. \\
\texttt{compute\_diagnostic\_indices} & Computes all ten diagnostic indices collectively.\\
\texttt{add\_group\_info} & Adds grouping information from the WDI data to the computed diagnostic indices. \\
\midrule
\texttt{plot\_metric\_distribution} & Visualises grouped or ungrouped distribution plots for all or selected diagnostic indices.\\
\texttt{plot\_metric\_partition} & Visualises metric values for each country, partitioned by the specified grouping variable.\\
\texttt{plot\_data\_trajectories} & Grouped or ungrouped interactive line plots of the series with or without metric-based highlighting. \\
\texttt{plot\_parallel\_coords} & Parallel coordinate plot of all the diagnostic indices. \\
\texttt{plot\_metric\_linkview} & Interactive linked view of the relationship between two metrics and their associated series trajectories by country.\\
\bottomrule
\end{tabular}
\end{table}

\subsection{Data functions}

The \texttt{get\_wdi\_data} function sources and locally stores WDI indicators datasets one at a time, leveraging the \textit{WDI} package. During this process, the indicator variable name is stored as a dataset attribute, which subsequent package functions use by default as the \texttt{index} argument ensuring consistency while avoiding the need to specify the variable with the country-year data points in any WDI data repeatedly. 

Country-level panel data often include missing observations and irregular time intervals; to address this, we extend the functionality of the \texttt{vis\_miss} function from the \textit{naniar} \citep{naniar} package by introducing the \texttt{plot\_missing} function that provides a visual summary of data availability and missingness across countries and years. Complementing this, the \texttt{get\_valid\_data} function inspects the dataset to identify countries and years with no valid data points. A data point is considered invalid if its indicator value is missing, that is $\mathrm{NA, NaN}$. The function returns a valid dataset with a printed summary of excluded countries and years.

\subsection{Diagnostic indices functions}

The functions that compute the collection of diagnostic indices, introduced in Section 3.1, are organised according to their respective categories: variation features, trend and shape features and sequential temporal features. Specifically, the \texttt{compute\_dissimilarity} function returns the dissimilarity matrix quantifying the Euclidean distance between countries computed with the \texttt{daisy()} function which handles missing values.
This dissimilarity matrix is used as an input to \texttt{compute\_variation}, which computes the average dissimilarity for each country, the average dissimilarities between a country and every other countries in its pre-defined group, and the silhouette width. 

The function \texttt{compute\_trend\_shape\_features} quantifies the directional patterns and shape of each series with trend strength, linearity, curvature and smoothness measures. The \texttt{compute\_temporal\_features} captures sequential temporal features including the number of crossing points, longest flat spot and autocorrelation. 

\texttt{compute\_diagnostic\_indices} computes all diagnostic indices collectively. \texttt{add\_group\_info} function adds the grouping information of the WDI data to the output of any function that computes diagnostic indices. This collection of diagnostic indices offers statistical measures that characterise the behavioural patterns as well as group-based patterns of country-level panel data. 

\subsection{Visualisation functions}

The remaining functions focus on visualising the computed diagnostic indices, providing both static and interactive visualisation functions. The \texttt{plot\_metric\_distribution} function generates a plot that displays the distribution of diagnostic indices. The function is flexible: it can create either an ungrouped or grouped plot version of the distribution for the values of a single metric, a selected subset of metric, or all the diagnostic indices collectively. When visualising a single metric, the function produces a straightforward distribution plot, while multiple metrics are displayed using faceting, each metric appears in its own panel, enabling clear visual comparison of their shapes and spreads. Within each panel, individual country-level metric values are represented by dots. See Figure \ref{fig:ungrouped-distribution-plot} for an example.

The \texttt{plot\_metric\_partition} function creates a layered bar plot that displays country-level metric values, grouped by a specified variable, with individual bars overlaid on lighter bars representing group-level averages. Each group level is distinguished by a unique colour, with the associated countries displayed accordingly, thereby facilitating comparison both within and across groups. Figure \ref{fig:partition-plot} is an example of a partition plot with the silhouette width metric.

Both \texttt{plot\_metric\_distribution} and \texttt{plot\_metric\_partition} functions generate static plots because the dots in the distribution plot are not rendered interactively, and the bars in the partition plot are sufficient to convey the metric values of each country, as tracing them toward the x-axis indicates their respective values.

Additionally, the package includes three interactive visualisation functions. The \texttt{plot\_data\_trajectories} function creates interactive line plots of the data series over time. It supports two modes: one that displays all series uniformly, and another that also displays the series uniformly while highlighting countries that fall within a specified threshold of the metric values. In both cases, the function uses interactive tooltips from the \textit{ggiraph} package to display the corresponding metric values. Each mode can be rendered in two versions: ungrouped and grouped. In the ungrouped version, all countries are plotted together without any grouping structure, and when highlighting is enabled, it is based on a global percentile threshold derived from the overall metric distribution. In the grouped version, countries are grouped by a specified variable; when highlighting is applied, the threshold is calculated within each group separately, allowing the plot to reflect group-level variation in the metric rather than the overall distribution. Figure \ref{fig:ungrouped-dissimilarity-plot} is an example of ungrouped data trajectories plot.

The \texttt{plot\_parallel\_coords} function generates two versions of interactive parallel coordinates plot, where each line corresponds to a country and spans the ten diagnostic indices. Tooltips appear when hovering over a line across the x-axis, displaying the corresponding country name, metric and its value for that country, facilitating a detailed comparison across all indices. The ungrouped version colour the lines by the specified grouping variable, while the grouped version also colour the lines by the specified grouping variable and facet the parallel coordinates plots by the specified grouping variable. Figure \ref{fig:ungrouped-parallel-plot} is an example of a parallel coordinate plot.

The \texttt{plot\_metric\_linkview} function offers an interactive dual-view display that links the relationship between two metric values to their corresponding series line plots, enabling exploration of how they relate to the overall data trajectories. See Figure \ref{fig:ungrouped-linkview-plot} for an example.

The majority of these functions accept an \texttt{index} argument that, by default, uses the first variable saved as the WDI dataset attribute. A consistent colour scheme is used across all visualisation functions to distinguish countries within the same group, facilitating comparison within and between groups. Additionally, we adopt a clear and descriptive naming format across all functions, aligned with their intended purpose and the overall structure of the package. These choices improve usability and support an intuitive workflow.

\hypertarget{workflow}{%
\section{Package Workflow}\label{workflow}}
In this section, we demonstrate the \textit{wdiexplorer} R package workflow categorised into three main stages, using the PM$_{2.5}$ air pollution data as an example. This data is derived from the Global Burden of Disease Study 2021 (GBD 2021), coordinated by the Institute for Health Metrics and Evaluation providing air pollution exposure estimates for nitrogen dioxide pollution, ozone pollution, ambient particulate matter pollution, and household air pollution. The PM$_{2.5}$ air pollution dataset contains mean annual exposure levels to ambient PM$_{2.5}$, measured in micrograms per cubic meter $(\mu g / m^3)$.

\subsection{Stage 1: Data Sourcing and Preparation}\label{stage1}
This initial stage of the workflow uses three core functions: \texttt{get\_wdi\_data}, \texttt{plot\_missing} and \texttt{get\_valid\_data}.

The \texttt{get\_wdi\_data} function sources the WDI indicator data of interest, taking a single argument \texttt{indicator}, which should be a valid WDI indicator code (e.g., PM$_{2.5}$ air pollution data with WDI indicator code: "EN.ATM.PM25.MC.M3"). To find any indicator code, users can employ the \texttt{WDI::WDISearch()} function with a relevant search keyword as its argument as illustrated below.

\begin{verbatim}
WDI::WDIsearch("air pollution")

pm_data <- get_wdi_data(indicator = "EN.ATM.PM25.MC.M3")
\end{verbatim}
The dataset for any specified indicator code is a data frame with 13 variables and is stored locally in a ``wdi\_data'' folder for easy future access. The 13 variables of any WDI indicator dataset are described in Table \ref{variables-table} below.

\begin{table}[H]
\centering
\caption{WDI variable names and descriptions.}
\label{variables-table}
\begin{tabular}{lp{0.75\linewidth}} 
\toprule
Variable name & Description \\
\midrule
  Country &  Country name. \\
  iso2c &  2-letter ISO country code. \\
  iso3c &  3-letter ISO country code. \\
  year &  Calendar year representing the time index of the observation. \\
  indicator code &  The variable containing values for the specified indicator code. ``In this package, this serves as the index variable''. \\
  status &  Describes the nature of the reported value, whether it is actual, estimated, or forecasted. \\
  lastupdated &  Timestamp that indicates the most recent update of the indicator values. \\
  region &  Geographical region. \\
  capital &  Name of the capital city. \\
  longitude &  Geographic coordinate that measures how far east or west a country is from the Prime Meridian. \\
  latitude &  Geographic coordinate that measures how far north or south a country is from the Equator. \\
  income &  World Bank income classification. \\
  lending &  World Bank lending classification. \\
\bottomrule
\end{tabular}
\end{table}
\begin{verbatim}
tibble::glimpse(pm_data)

 Downloading WDI indicator: EN.ATM.PM25.MC.M3 data using the WDI R package.
Rows: 13,910
Columns: 13
$ country           <chr> "Afghanistan", "Afghanistan", "Afghanistan"…
$ iso2c             <chr> "AF", "AF", "AF", "AF", "AF", "AF", "AF", "…
$ iso3c             <chr> "AFG", "AFG", "AFG", "AFG", "AFG", "AFG", "…
$ year              <int> 1964, 1965, 2007, 1963, 1970, 1969, 1968, 1…
$ EN.ATM.PM25.MC.M3 <dbl> NA, NA, 58.08379, NA, NA, NA, NA, NA, NA, 5…
$ status            <chr> "", "", "", "", "", "", "", "", "", "", "",…
$ lastupdated       <chr> "2025-07-01", "2025-07-01", "2025-07-01", "…
$ region            <chr> "South Asia", "South Asia", "South Asia", "…
$ capital           <chr> "Kabul", "Kabul", "Kabul", "Kabul", "Kabul"…
$ longitude         <chr> "69.1761", "69.1761", "69.1761", "69.1761",…
$ latitude          <chr> "34.5228", "34.5228", "34.5228", "34.5228",…
$ income            <chr> "Low income", "Low income", "Low income", "…
$ lending           <chr> "IDA", "IDA", "IDA", "IDA", "IDA", "IDA", "…
\end{verbatim}

Once the data is sourced, an important initial step is to highlight missing entries across countries and time periods. To facilitate this, the \texttt{plot\_missing} function takes two arguments: a dataset containing data for a selected WDI indicator, and a pre-defined grouping variable within the dataset. It also includes an additional argument \texttt{index}, which defaults to \texttt{NULL} and uses the first variable saved as the WDI dataset attribute. The function produces a missingness plot that provides a structured overview of missing data and shows its distribution over time, across countries, and by the specified grouping variable. 

\begin{figure}[H]
  \centering
  \includegraphics[width=\textwidth, height=0.8\textheight, keepaspectratio]{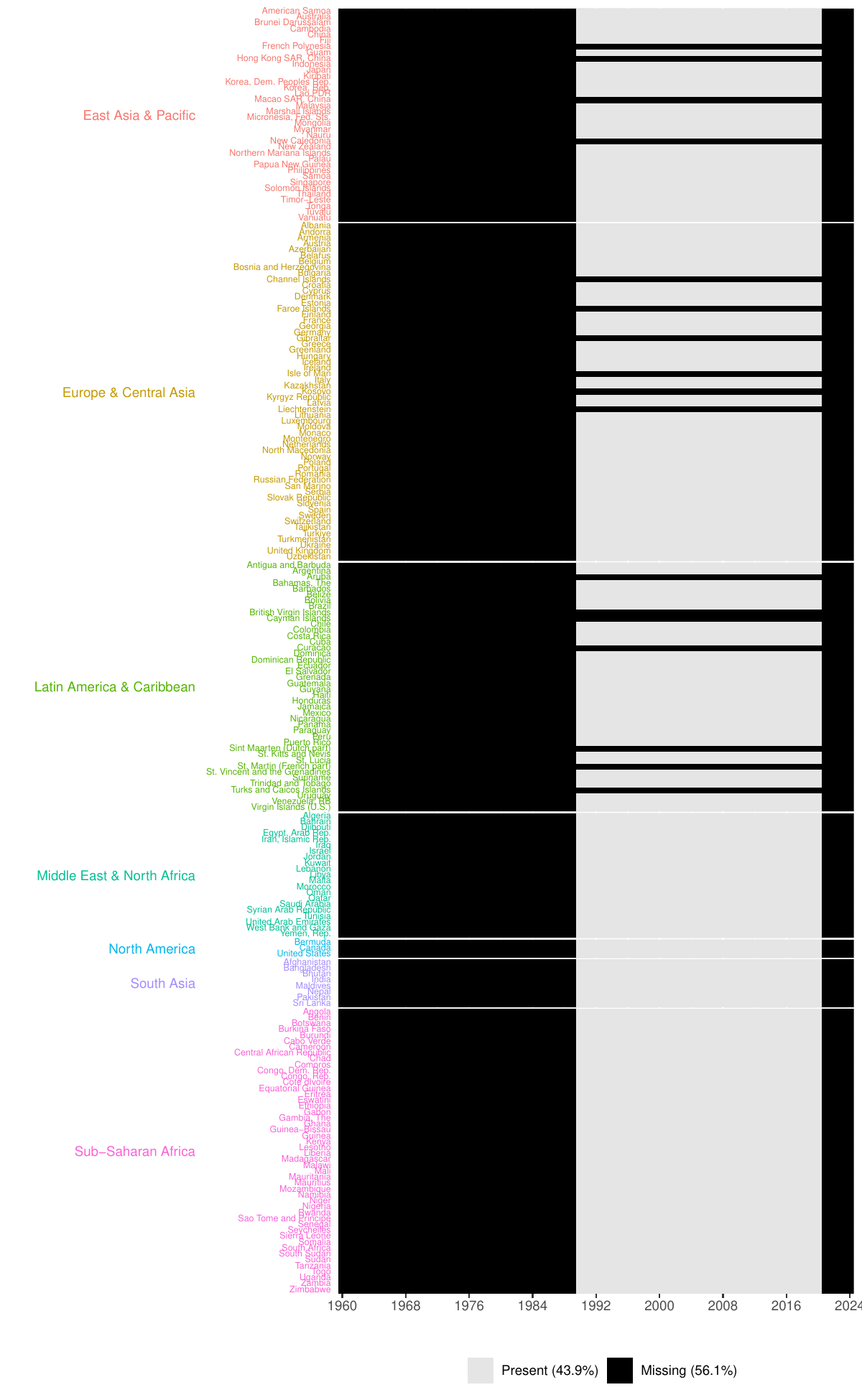}
  \caption{Missingness plot, providing information about the years and countries with missing entries and the overall percentages of missing and present data. Some countries in East Asia \& Pacific, Europe \& Central Asia, and Latin America \& Caribbean regions have no recorded PM$_{2.5}$ air pollution data. It also shows that no data points are available in any country during the years 1960 to 1989 and 2021 to 2024.}
  \label{fig:missingness-plot}
\end{figure}

The missingness plot in Figure \ref{fig:missingness-plot} shows that no data are present from 1960 to 1989 and from 2021 to 2024. During the period 1990 to 2020, data are available across all countries in the Middle East \& North Africa, South Asia, and Sub-Saharan Africa. In contrast, several countries in other regions have no valid data recorded. Specifically, four countries in East Asia \& Pacific, six countries in Europe \& Central Asia, and seven countries in Latin America \& the Caribbean. In total, 17 countries across these three regions have no valid data between 1990 and 2020. These patterns of missingness are clearly captured by the plot, which provides a structured overview across time, countries, and pre-defined groups (geographic or economic groupings).

\begin{verbatim}
    plot_missing(pm_data, group_var = "region")
\end{verbatim}

To complement this visual summary, we introduce a second step: calculating the total number of missing entries per country. This quantitative assessment mirrors the information shown in the missingness plot and helps identify countries with particularly high levels of missing data within their group. The following code demonstrates this using our illustrative dataset.

\begin{verbatim}
index <- "EN.ATM.PM25.MC.M3"

pm_data |>
  dplyr::select(country, region, year, tidyselect::all_of(index)) |>
  dplyr::group_by(region, country) |>
  naniar::miss_var_summary() |>
  dplyr::filter(variable == index) |>
  dplyr::arrange(desc(n_miss))

# A tibble: 214 × 5
# Groups:   region, country [214]
   country                region                    variable          n_miss pct_miss
   <chr>                  <chr>                     <chr>              <int>    <num>
 1 Aruba                  Latin America & Caribbean EN.ATM.PM25.MC.M3     65      100
 2 British Virgin Islands Latin America & Caribbean EN.ATM.PM25.MC.M3     65      100
 3 Cayman Islands         Latin America & Caribbean EN.ATM.PM25.MC.M3     65      100
 4 Channel Islands        Europe & Central Asia     EN.ATM.PM25.MC.M3     65      100
 5 Curacao                Latin America & Caribbean EN.ATM.PM25.MC.M3     65      100
 6 Faroe Islands          Europe & Central Asia     EN.ATM.PM25.MC.M3     65      100
 7 French Polynesia       East Asia & Pacific       EN.ATM.PM25.MC.M3     65      100
 8 Gibraltar              Europe & Central Asia     EN.ATM.PM25.MC.M3     65      100
 9 Hong Kong SAR, China   East Asia & Pacific       EN.ATM.PM25.MC.M3     65      100
10 Isle of Man            Europe & Central Asia     EN.ATM.PM25.MC.M3     65      100
# 204 more rows
\end{verbatim}

In addition, the \textit{wdiexplorer} package provides the \texttt{get\_valid\_data} function, which reports countries with no data points as well as years for which no data are available, and returns a tibble with the valid data for the provided WDI indicator dataset. The function takes one argument, a dataset containing data for a selected WDI indicator.

\begin{verbatim}
get_valid_data(pm_data)

The 17 countries listed below had no available data and were excluded:
 Aruba
- British Virgin Islands
- Cayman Islands
- Channel Islands
- Curacao
- Faroe Islands
- French Polynesia
- Gibraltar
- Hong Kong SAR, China
- Isle of Man
- Kosovo
- Liechtenstein
- Macao SAR, China
- New Caledonia
- Sint Maarten (Dutch part)
- St. Martin (French part)
- Turks and Caicos Islands

The 34 year(s) listed below had no available data and were excluded:
 1960, 	1961, 	1962, 	1963, 	1964, 	1965, 	
 1966, 	1967, 	1968, 	1969, 	1970, 	1971, 	
 1972, 	1973, 	1974, 	1975, 	1976, 	1977, 	
 1978, 	1979, 	1980, 	1981, 	1982, 	1983, 	
 1984, 	1985, 	1986, 	1987, 	1988, 	1989, 	
 2021, 	2022, 	2023, 	2024 

\end{verbatim}

The \texttt{get\_valid\_data} function is primarily used to investigate missing entries and identify countries excluded from the indicator data. This function is executed internally by all other functions in the \textit{wdiexplorer} workflow to filter valid data from the specified indicator data.

\subsection{Stage 2: Diagnostic Indices}\label{indices} 

This second stage of the workflow focuses on calculating the diagnostic indices introduced in Section 3.1. They measure variation, trend and shape features, as well as sequential temporal characteristics. There are three functions in the \textit{wdiexplorer} package which support the computation of these indices. 

\subsubsection{Variation Features}
To measure variation, the function \texttt{compute\_dissimilarity} takes one main argument, a dataset of any WDI indicator, and returns a matrix of dissimilarity values between country pairs. The function also includes two additional arguments with default values: \texttt{index}, which defaults to \texttt{NULL} and uses the first variable saved as the WDI dataset attribute; and \texttt{metric}, which specifies the dissimilarity calculation method and defaults to Euclidean distance.

The function \texttt{compute\_variation} accepts two main arguments: a dataset of any WDI indicator and a grouping variable \texttt{group\_var}. It also includes an optional dissimilarity matrix argument, \texttt{diss\_matrix} (defaulting to the output of \texttt{compute\_dissimilarity}. 
The \texttt{compute\_variation} function returns a data frame with \texttt{country}, \texttt{country\_avg\_dist}, average distance between each country and all other countries in the data set, \texttt{within\_group\_avg\_dist}, average distance between each country and all other countries within its pre-defined group, and \texttt{sil\_width}, silhouette widths which measure how well each country fits within its group compared to other groups. The code below demonstrates the use of these functions:

Users can compute the dissimilarity matrix separately and pass it directly as the \texttt{diss\_matrix} argument into the \texttt{compute\_variation} function as demonstrated below or allow the function to compute it internally by specifying only the two main arguments:

\begin{verbatim}
pm_diss_mat <- compute_dissimilarity(pm_data)
 
pm_variation <- compute_variation(
                    pm_data, 
                    diss_matrix = pm_diss_mat, 
                    group_var = "region"
        )
\end{verbatim}

The output \texttt{pm\_variation} enables the exploration of computed variation features. It facilitates the identification of the most distinctive countries, the evaluation of within-group differences, and the analysis of how closely aligned countries within a group are compared to those in other groups. However, these measures are not always intuitive to interpret on their own; they are best understood in conjunction with the accompanying line plot, which displays the behavioural patterns over time, offering meaningful context for comparing the trajectories of standout countries to the rest.

\begin{verbatim}
# country dissimilarity average
pm_variation |> 
        dplyr::arrange(desc(country_avg_dist)) |> 
        dplyr::slice_head(n = 3)
        
# A tibble: 3 × 5
  country    group            country_avg_dist within_group_avg_dist sil_width
  <chr>      <chr>                       <dbl>                 <dbl>     <dbl>
1 Qatar      Middle East & N…             498.                  410.    -0.168
2 Niger      Sub-Saharan Afr…             479.                  408.    -0.207
3 Mauritania Sub-Saharan Afr…             415.                  345.    -0.243

# within group dissimilarity average
 pm_variation |> 
        dplyr::arrange(desc(within_group_avg_dist)) |> 
        dplyr::slice_head(n = 3)
 
# A tibble: 3 × 5
  country    group            country_avg_dist within_group_avg_dist sil_width
  <chr>      <chr>                       <dbl>                 <dbl>     <dbl>
1 Qatar      Middle East & N…             498.                  410.    -0.168
2 Niger      Sub-Saharan Afr…             479.                  408.    -0.207
3 Mauritania Sub-Saharan Afr…             415.                  345.    -0.243
\end{verbatim}

The results above show that Qatar has the highest overall average dissimilarity, followed by Niger and Mauritania. These countries, Qatar, Niger, and Mauritania, not only rank highest in overall average dissimilarity but also exhibit the highest dissimilarity relative to countries in their respective region groups. 

\begin{verbatim}
pm_variation |> 
    dplyr::arrange(desc(sil_width)) |>
    dplyr::slice_head(n = 3)

# A tibble: 3 × 5
  country       group         country_avg_dist within_group_avg_dist sil_width
  <chr>         <chr>                    <dbl>                 <dbl>     <dbl>
1 Canada        North America             151.                  13.7     0.836
2 Bermuda       North America             155.                  17.2     0.798
3 United States North America             138.                  23.0     0.695
\end{verbatim}
As shown in the \texttt{pm\_variation} result, Canada, Bermuda, and the United States (the only countries from the North America region group with available PM$_{2.5}$ air pollution data) exhibit the highest silhouette widths, indicating that these countries have similar PM$_{2.5}$ trajectories, which are dissimilar to those in other region groups.

\subsubsection{Trend and Shape Features}

To examine trend and shape features, we implement \texttt{compute\_trend\_shape\_features}, which takes as the main argument a dataset of any WDI indicator. An additional argument \texttt{index} specifies the indicator name. If \texttt{index} is \texttt{NULL} then the first variable saved as the WDI dataset attribute is used. The function returns a data frame with one row per country and columns for the  measures \texttt{trend strength}, \texttt{linearity}, \texttt{curvature}, and \texttt{smoothness}.

\begin{verbatim}
pm_trend_shape <- compute_trend_shape_features(pm_data)
\end{verbatim}

The \texttt{pm\_trend\_shape} output enables the exploration of the computed trend and shape features. 

\begin{verbatim}
pm_trend_shape |> 
        dplyr::arrange(desc(trend_strength)) |>
    dplyr::slice_head(n = 3)

# A tibble: 3 × 5
  country        trend_strength linearity curvature smoothness
  <chr>                   <dbl>     <dbl>     <dbl>      <dbl>
1 Ukraine                 0.996     -32.4     2.69       0.547
2 Moldova                 0.995     -34.4     2.65       0.666
3 United Kingdom          0.995     -16.3    -0.217      0.329
\end{verbatim}

The output above shows that Ukraine, Moldova, and the United Kingdom are the three countries with the strongest trend strength. In this context, trend strength measures the extent to which data follows a consistent pattern over time, whether linear or curved.

\begin{verbatim}
pm_trend_shape |> 
     dplyr::arrange(desc(linearity)) |>
     dplyr::slice(c(1, 2, dplyr::n(), dplyr::n() - 1))

# A tibble: 4 × 5
  country      trend_strength linearity curvature smoothness
  <chr>                 <dbl>     <dbl>     <dbl>      <dbl>
1 Mongolia              0.944      27.2      4.47       1.95
2 Saudi Arabia          0.810      25.5    -11.7        4.34
3 Bolivia               0.995    -116.       2.46       2.08
4 Peru                  0.991    -104.     -10.2        2.49
\end{verbatim}

The output above lists countries with the strongest positive and negative linear trend as represented by the \texttt{linearity column}, Mongolia and Saudi Arabia exhibit the highest positive linearity values, and  Bolivia and Peru are countries with the highest negative linearity. A more detailed interpretation of these results is presented in Section 5.3, where the metric values are explored in conjunction with the trajectories of each country’s data series.

The next code segment identifies countries with extreme curvature values.
\begin{verbatim}
pm_trend_shape |> 
    dplyr::arrange(desc(curvature)) |>
     dplyr::slice(c(1, 2, dplyr::n(), dplyr::n() - 1))

# A tibble: 4 × 5
  country   trend_strength linearity curvature smoothness
  <chr>              <dbl>     <dbl>     <dbl>      <dbl>
1 Singapore          0.970    -21.1       14.4       1.41
2 Senegal            0.575    -13.7       10.1       6.19
3 Kuwait             0.762     13.7      -17.2       4.92
4 Libya              0.713      6.78     -15.2       3.94
\end{verbatim}
From the above output, Singapore and Senegal exhibit the most positive curvature values, indicating that their series follow a U-shaped pattern. Conversely, Kuwait and Libya exhibit the most negative curvature values, indicating a pronounced inverted U-shaped pattern. 

We proceeded to identify countries with the highest and lowest smoothness values.
\begin{verbatim}
pm_trend_shape |> 
    dplyr::arrange(desc(smoothness)) |> 
    dplyr::slice(c(1, dplyr::n()))

# A tibble: 2 × 5
  country trend_strength linearity curvature smoothness
  <chr>            <dbl>     <dbl>     <dbl>      <dbl>
1 Niger            0.242    13.9      -9.88       8.54 
2 Tuvalu           0.906     0.631    -0.596      0.144
\end{verbatim}
This output shows that Niger, a country in Middle East \& North Africa region, emerges as the country with the highest smoothness value. Tuvalu, located in the East Asia \& Pacific region, has the lowest smoothness value.

\subsubsection{Sequential Temporal Features}
Lastly, to measure the sequential temporal features of the data series, we implement a function named  \texttt{compute\_temporal\_features}, which takes the same arguments as the other functions that calculate diagnostic indices. The function returns a data frame where each row represents a country, and columns for the \texttt{country} name, and measures of \texttt{number of crossing points}, \texttt{longest flat spot}, and \texttt{autocorrelation}.

\begin{verbatim}
pm_temporal <- compute_temporal_features(pm_data)
\end{verbatim}

We can use \texttt{pm\_temporal} to explore of the computed sequential temporal features.

\begin{verbatim}
pm_temporal |> 
   dplyr::arrange(desc(crossing_points)) |> 
   dplyr::slice(c(1, 2, dplyr::n(), dplyr::n() - 1))

# A tibble: 4 × 4
  country       crossing_points flat_spot   acf
  <chr>                   <int>     <int> <dbl>
1 Gabon                      10         4 0.338
2 Tunisia                    10         5 0.760
3 Venezuela, RB               1        13 0.942
4 Uruguay                     1        11 0.977
\end{verbatim}
The code segment above shows that Gabon and Tunisia have the highest number of crossing points. In contrast, 54 countries have only one crossing point, indicating that these countries cross above or below their median value just once.

We proceeded to identify countries with the highest and the lowest number of flat spot: 
\begin{verbatim}
pm_temporal |> 
   dplyr::arrange(desc(flat_spot)) |> 
   dplyr::slice(c(1:3, (dplyr::n() - 2):dplyr::n()))

# A tibble: 6 × 4
  country              crossing_points flat_spot      acf
  <chr>                          <int>     <int>    <dbl>
1 Kyrgyz Republic                    3        18  0.909  
2 United Arab Emirates               9        18 -0.00320
3 Armenia                            1        17  0.934  
4 Jordan                             3         3  0.846  
5 Rwanda                             7         3  0.539  
6 West Bank and Gaza                 8         3  0.608  
\end{verbatim}
We see from this output that Kyrgyz Republic and United Arab Emirates have the longest flat spots, characterised by long consecutive periods during which their data series remain within an interval. In contrast, Jordan, Rwanda, and West Bank and Gaza exhibit the shortest consecutive periods where their series remain within a specified interval.

The next code segment investigates the acf measures in \texttt{pm\_temporal}:
\begin{verbatim}
pm_temporal |> 
     dplyr::arrange(desc(acf)) |>
     dplyr::slice(c(1, dplyr::n()))
     
# A tibble: 2 × 4
  country              crossing_points flat_spot      acf
  <chr>                          <int>     <int>    <dbl>
1 Ukraine                            1         5  0.996  
2 United Arab Emirates               9        18 -0.00320
\end{verbatim}
The above output shows that Ukraine has the highest autocorrelation value. United Arab Emirates has  zero autocorrelation value. United Arab Emirates has the zero autocorrelation value and is also one of the countries with the longest flat spot. This suggests that although the data series in United Arab Emirates contains extended periods of minimal change reflected in its high number of longest flat spot, the zero autocorrelation indicates an absence of dependence between successive time points, even when its values remain within the same interval for the longest period. This shows that the diagnostic measures are often not informative enough when interpreted in isolation; they become more insightful when combined with their corresponding series trajectories as shown in Sub-section 5.3.

The \texttt{compute\_diagnostic\_indices} function returns all diagnostic indices collectively in a single data frame. The function takes two arguments: a dataset of any WDI indicator data and a grouping variable \texttt{group\_var} and returns a data frame with: \texttt{country}; \texttt{country\_avg\_dist}, \texttt{within\_group\_avg\_dist}, \texttt{sil\_width}, \texttt{trend strength}, \texttt{linearity}, \texttt{curvature}, \texttt{smoothness}, \texttt{number of crossing points}, \texttt{longest flat spot}, and \texttt{autocorrelation} measures. 

\begin{verbatim}
pm_diagnostic_metrics <- compute_diagnostic_indices(pm_data, group_var = "region")
\end{verbatim}

This \texttt{pm\_diagnostic\_metrics} output can be passed directly to the plot functions to generate both static and interactive visuals of the \textit{wdiexplorer} package.

The \texttt{add\_group\_info} function appends the pre-defined grouping information from the WDI dataset to the data frame of any computed diagnostics function output. The function takes two arguments: a data frame with the calculated diagnostic indices \texttt{metric\_summary}; and a dataset of any WDI indicator data. This facilitates the direct use of the resulting data frame with the plot functions to create group versions of the visualisations.

\begin{verbatim}
pm_diagnostic_metrics_group <- add_group_info(
                    metric_summary = pm_diagnostic_metrics,
                    pm_data
            )
\end{verbatim}

\subsection{Stage 3: Static and Interactive Visualisations}\label{visuals}

The third stage of the workflow utilises visual summaries to detect potentially interesting features within panel data. Our package offers five core functions, two static plot functions: \texttt{plot\_metric\_distribution}, and \texttt{plot\_metric\_partition} and three interactive plot functions:
\texttt{plot\_data\_trajectories}, \texttt{plot\_parallel\_coords}, and \texttt{plot\_metric\_linkview}, all built on the \textit{ggplot2} \citep{ggplot2} framework and using \textit{ggiraph} \citep{ggiraph} for interactivity, \textit{ggdist} \citep{ggdist} for the distribution plots and \textit{scales} \citep{scales} for managing pre-defined group colour palettes. Each of these functions is demonstrated below.

\subsubsection{Distribution Plot}
The \texttt{plot\_metric\_distribution} function takes two main arguments: a data frame containing the computed diagnostic indices and the pre-defined grouping information \texttt{metric\_summary}; and a variable, \texttt{colour\_var} whose levels are mapped to distinct colours in the resulting dot plot. An optional argument, \texttt{group\_var}, controls whether the distribution is grouped or ungrouped. To generate the distribution plot for specific metrics, users can pass the variable names to the \texttt{metric\_var} argument. 

The code below demonstrates how to use the \texttt{plot\_metric\_distribution} function to generate ungrouped distribution plots for all diagnostic indices.

\begin{verbatim}
plot_metric_distribution(
      metric_summary = pm_diagnostic_metrics_group, 
      colour_var = "region"
      )
\end{verbatim}

\begin{figure}[H]
  \centering
  \includegraphics[width=\textwidth]{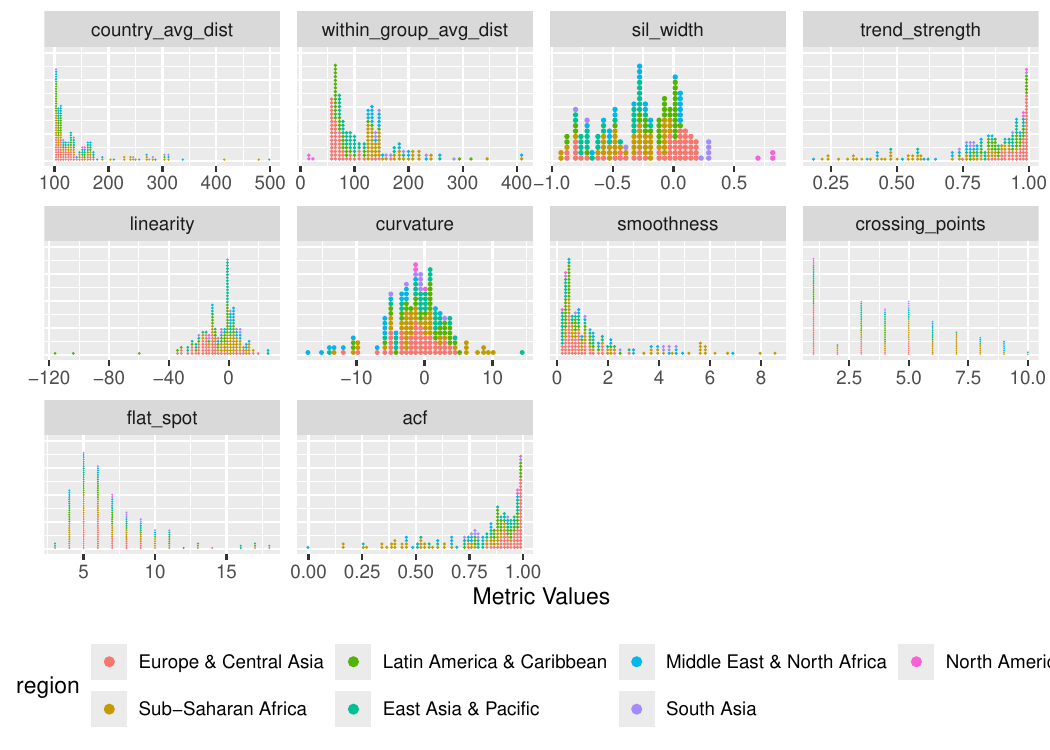}
  \caption{Distribution of diagnostic indices where each panel represents a different metric. It shows the spread of the metric values across countries, with each dot representing a country and coloured by region. Countries in the North America region stand out with the lowest within-group average dissimilarity and the highest silhouette width values.}
  \label{fig:ungrouped-distribution-plot}
\end{figure}

Figure \ref{fig:ungrouped-distribution-plot} shows the distribution of all diagnostic indices using dotplots. Each metric is presented in a separate panel, with each dots showing the per country metric value with dots coloured by region. The figure reveals distinct distributional patterns across the indices. For instance, country average dissimilarity and smoothness measures are right skewed with most countries having low values. These indicate that the majority of countries differ only minimally from one another and tend to follow smooth, gradual changes over time (based on the smoothness measures). In contrast, trend strength and autocorrelation (acf) are left skewed, with most countries showing high metrics values. These reveal that, in majority of the countries, their data series exhibit persistence trends which could be linear or curved and high correlation between successive time points. However, linearity and curvature are centred around zero. Linearity ($\beta_1$) value close to $0$ suggests the absence of any linear trend and curvature ($\beta_2$) value close to $0$ indicates that the trend follows a nearly linear trajectory, with minimal or no curvature. Although the plot is not interactive and the countries corresponding to the dots centred around zero cannot be identified, a literal interpretation suggests that the trends in some countries are neither predominantly linear nor strongly curved.

The Figure \ref{fig:ungrouped-distribution-plot} also highlights that the three countries in the North America region shown in {\color{magenta}magenta} stand out with the lowest within-group average distance and the highest silhouette width values. The low within-group average distance indicates minimal variation among the temporal patterns of these countries, suggesting that their PM$_{2.5}$ air pollution trends are highly similar. The high silhouette width further confirms this, as it measures how well each country fits within its group relative to other groups, higher values indicate that these countries are closely aligned to their pre-defined group and distinct from others. 

The code below demonstrates how to use the \texttt{plot\_metric\_distribution} function to generate grouped distribution plots for all diagnostic indices.

\begin{verbatim}
plot_metric_distribution(
      metric_summary = pm_diagnostic_metrics_group, 
      colour_var = "region",
      group_var = "region"
      )
\end{verbatim}

\begin{figure}[H]
  \centering
  \includegraphics[width=\textwidth]{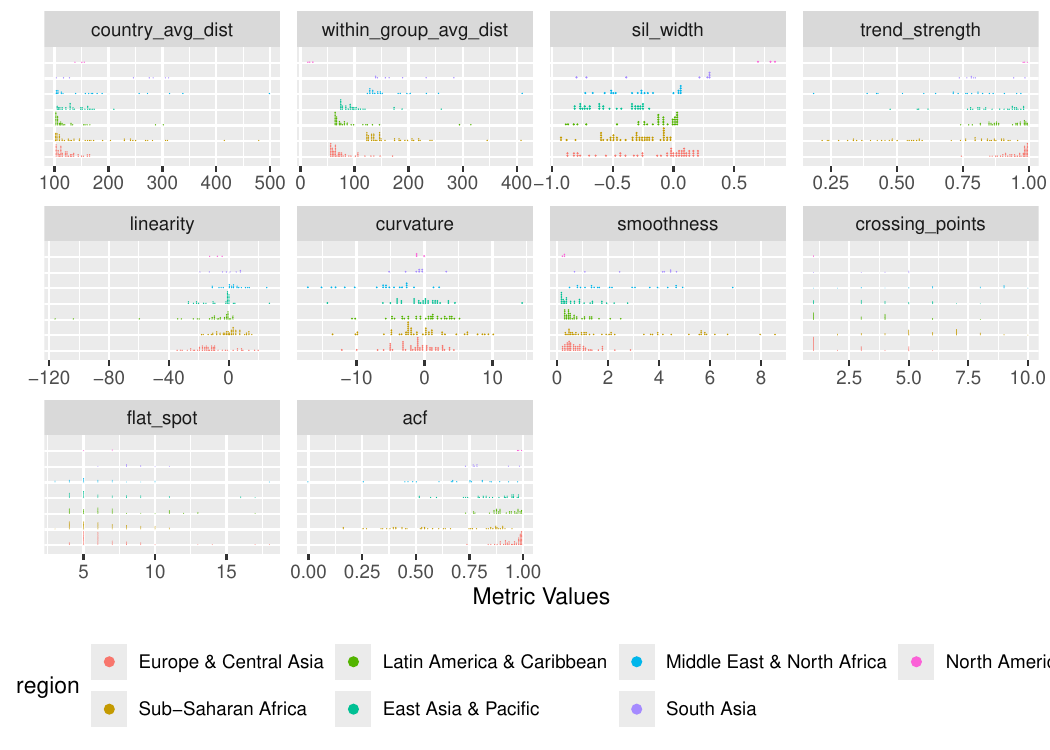}
  \caption{Distribution of diagnostic indices grouped by region. Each panel displays a metric, with countries organised by region to facilitate within and between group comparisons. The plot reveals region-specific patterns and outliers. Sub-Saharan Africa and East Asia \& Pacific regions show wider spread while North America region are closely aligned.}
  \label{fig:grouped-distribution-plot}
\end{figure}

Figure \ref{fig:grouped-distribution-plot} shows the distribution of diagnostic indices grouped by region. As in the ungrouped distribution plot in Figure \ref{fig:ungrouped-distribution-plot}, each panel represents a different metric. However, while the ungrouped version presents all countries together as individual dots in a single distribution per metric, the grouped version organises countries by region, making it easier to compare both within and between group metric values across regions. Across all regions, the country average dissimilarity metric is consistently right-skewed, though Sub-Saharan Africa and Middle East \& North Africa contain notable outliers that deviate substantially from other countries in their region. A similar pattern is observed in their within-group average dissimilarity metric. In the silhouette width panel, All countries in Latin America \& Caribbean and Sub-Saharan Africa have negative values, suggesting that their temporal patterns may be more aligned with countries outside their assigned groups while the three countries in the North America region have the highest silhouette width values. The North America region exhibit closely aligned metric values across all diagnostic indices with minimal differences. This indicates a highly knitted region with similar temporal behaviour.

Figure \ref{fig:grouped-distribution-plot} also highlights outlier countries in South Asia region where some of the countries exhibit negative silhouette width far off from the behaviour in other countries. Outliers are easily identified in most of the regions across metrics in the grouped distribution plot. For example, East Asia \& Pacific has the country with the highest positive curvature value while Middle East \& North Africa has countries with the most extreme negative curvature values. In smoothness metric, Sub-Saharan Africa has the highest smoothness values. Sub-Saharan Africa and Middle East \& North Africa exhibit broader spreads across metrics particularly in country average dissimilarity, trend strength, curvature, smoothness and autocorrelation (acf), reflecting substantial heterogeneity within the data series in these regions.

Although linearity and curvature remain centred around zero in most regions as previously seen in Figure \ref{fig:ungrouped-distribution-plot}, the ungrouped version and the grouped version in Figure \ref{fig:grouped-distribution-plot} reveals region-specific behaviours. For instance, Latin America \& Caribbean appears more left-skewed in linearity, indicating a pronounced downward trend in most of the countries.

\subsubsection{Partition Plot}

The \texttt{plot\_metric\_partition} function takes three arguments: \texttt{metric\_summary}, a data frame with the calculated diagnostic indices and the grouping information, \texttt{metric\_var}, a variable within the data frame that contains the metric values, and \texttt{group\_var} naming the grouping variable. The \texttt{metric\_summary} is the output of any diagnostic indices functions merged with the grouping information from the WDI data. This integration is performed using the \texttt{add\_group\_info} function, which appends the relevant group information of each country to the metric summary. The following code demonstrates \texttt{plot\_metric\_partition} using the silhouette width measures of the PM$_{2.5}$ air pollution data.

\begin{verbatim}
plot_metric_partition(
          metric_summary = pm_diagnostic_metrics_group,
          metric_var = "sil_width",
          group_var = "region"
 )
\end{verbatim}

\begin{figure}[H]
  \centering
  \includegraphics[width=\textwidth, height=0.93\textheight, keepaspectratio]{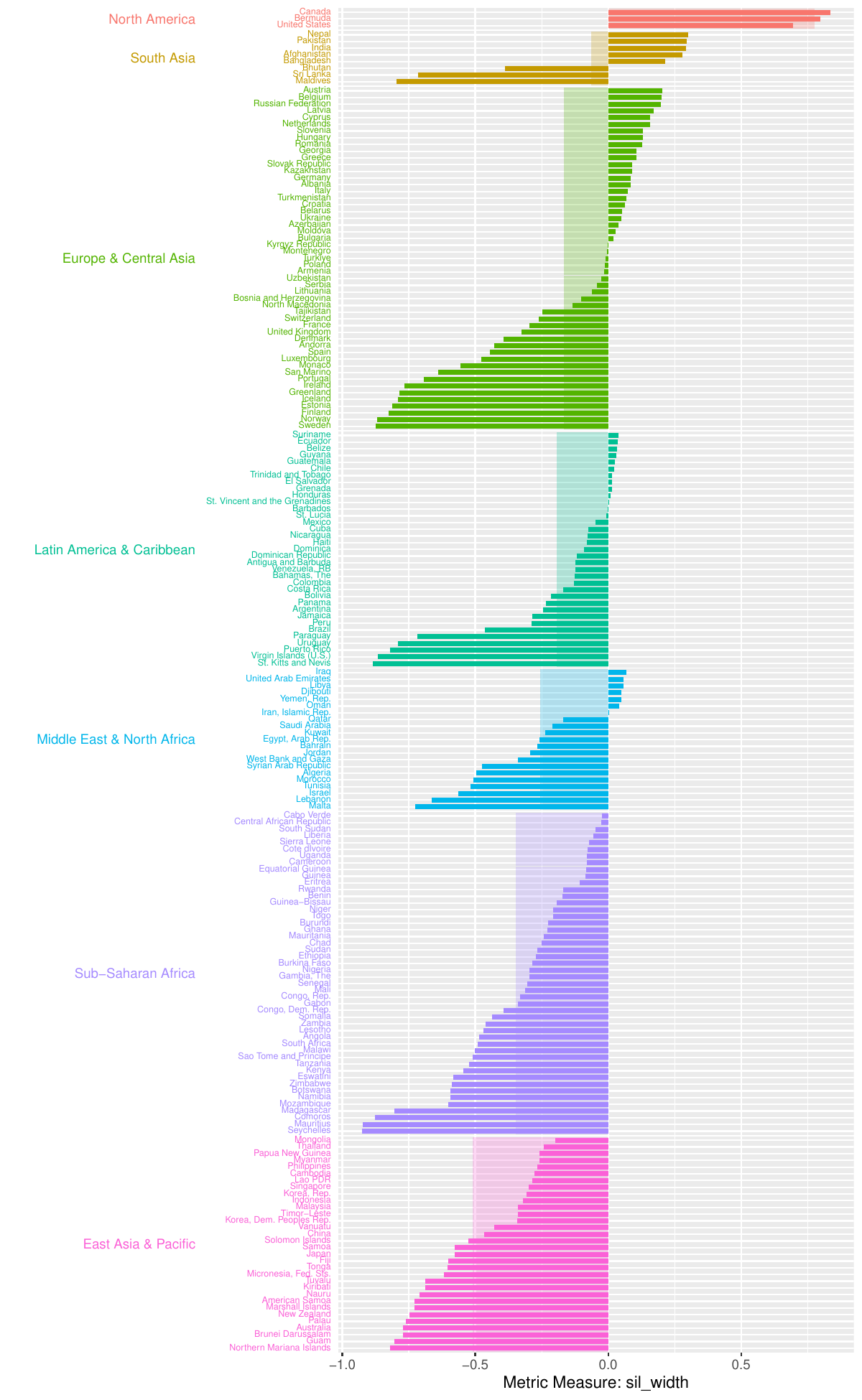}
  \caption{Country silhouette widths, grouped by region, with the average silhouette width for each region underlaid beneath the country bars. Countries in Sub-Saharan Africa and East Asia \& Pacific regions all exhibit negative silhouette widths, suggesting that they do not fit well within their pre-defined regional groupings based on their data series, or that their behaviour may be more similar to countries in other regions.}
  \label{fig:partition-plot}
\end{figure}

Figure \ref{fig:partition-plot} presents the silhouette width value for each country, partitioned by their respective region groupings.  Each bar represent the silhouette width of a country, while the group-level average is underlaid beneath the country bars. A distinct colour palette is used to differentiate region levels, with countries within the same group sharing the same colour.

North America stands out as the region with the most closely aligned data series, with Bermuda, Canada and the United States all exhibiting high positive close to $\mathrm{+}1$ silhouette widths. This indicates that these countries behave consistently in terms of their PM$_{2.5}$ air pollution trends and differ clearly from countries in other regions. In contrast, South Asia displays a mix of both positive and negative silhouette values. Notably, three countries in the region exhibit highly negative values, indicating a weaker fit within the South Asia group. A similar variation is observed in Latin America \& Caribbean, Europe \& Central Asia and Middle East \& North Africa, where a few countries in these regions have slightly positive values, the majority exhibit extreme negative values close to $\mathrm{-}1$ and some close to $0$ values reflecting weak within-group similarity among countries within these regions.

Sub-Saharan Africa and East Asia \& Pacific regions are characterised by uniformly negative silhouette widths, though with varying degree as some countries exhibit close to $\mathrm{-}1$, while some are close to $0$. This reflects noticeable differences in behaviour among countries within the same region. Figure \ref{fig:partition-plot} reveals that, only countries in North America exhibit consistently similar temporal patterns, while other regions exhibit weaker similarity with more group-wise variability.

The remaining visualisation functions are designed to produce interactive plots with tooltips displaying the country name and their respective metric value. For the purposes of this documentation, we provide static versions instead, while the interactive versions are available in an online archive via \url{https://oluwayomi-olaitan.github.io/rjournal-wdiexplorer-interactive-plots/}. The code examples below demonstrate how each of these functions can be used to interpret the diagnostic measures discussed in Section 5.2 in the data trajectories space.

\subsubsection{Data Trajectories Plot}

The \texttt{plot\_data\_trajectories} function generates interactive line plots. It takes one main argument: a dataset containing data for a selected WDI indicator. It also accepts an optional \texttt{index} argument which defaults to $\mathrm{NULL}$ and uses the first variable saved as the WDI dataset attribute. By default, it generates an ungrouped interactive line plot showing country-level trajectories. If an additional optional argument \texttt{group\_var} is specified, the function instead produces a grouped version of the interactive line plots of the series, faceted by the specified grouping variable. Additional arguments \texttt{metric\_summary} and \texttt{metric\_var} can be supplied to highlight the trajectories based on the specified percentile of the metric values. \texttt{metric\_summary} is a data frame containing the calculated diagnostic indices alongside the grouping information, while \texttt{metric\_var} specifies which variable in that data frame contains the values of the metric for highlighting.

The function also accepts a \texttt{percentile} argument, which defines the threshold for highlighting countries based on their metric values. By default, \texttt{percentile = $0.95$}, meaning the top $5\%$ of countries based on the selected metric are interactively emphasised with distinct colour palette. When \texttt{group\_var} is $\mathrm{NULL}$, the function highlights countries using a global threshold across all countries. When \texttt{group\_var} is specified, the function produces a faceted plot grouped by the specified variable, and countries are highlighted based on group-specific percentile thresholds rather than a global distribution threshold.

To demonstrate the functionality of the \texttt{plot\_data\_trajectories} for highlighting countries based on the computed metrics, we select two diagnostic indices: dissimilarity and linearity. We use absolute values of the linearity index to ensure that both strong positive and strong negative are treated equally when highlighting the top percentile. 

The ungrouped and grouped data series trajectories with the top 5\% countries highlighted based on their dissimilarities.
 
\begin{verbatim}
# ungrouped version
plot_data_trajectories(
        pm_data, 
        metric_summary = pm_diagnostic_metrics, 
        metric_var = "country_avg_dist"
    )
\end{verbatim}

\begin{figure}[H]
  \centering
  \includegraphics[width=\textwidth]{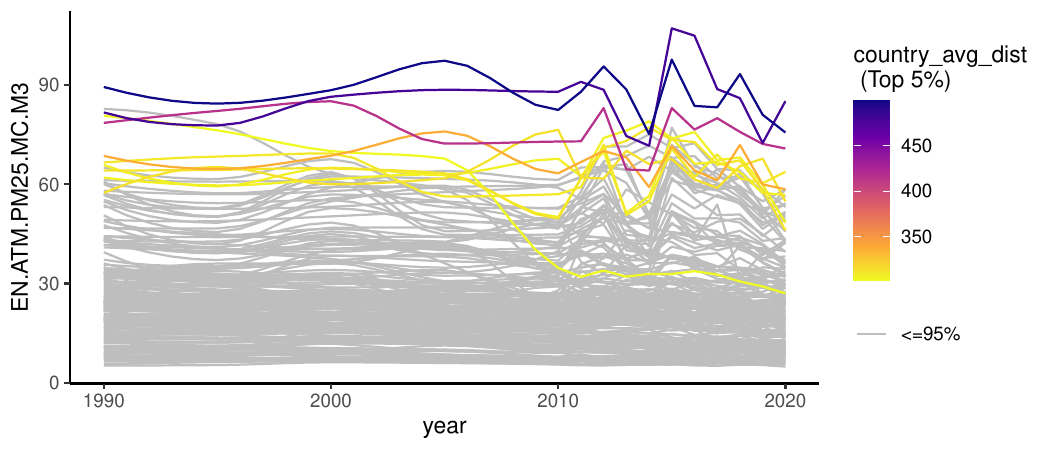}
  \caption{Country line plots of PM$_{2.5}$ air pollution dataset. Countries with average dissimilarity distance values below or at the 95th percentile are displayed in grey, while countries with the top 5\% average dissimilarity between itself and other countries are highlighted using a colour gradient. Qatar and Niger, countries displayed in {\color{purple-blue} purple-blue} exhibit the highest dissimilarity values. Hovering any of the highlighted lines in the \href{https://oluwayomi-olaitan.github.io/rjournal-wdiexplorer-interactive-plots/interactive-plots/ungrouped_dissimilarity_plot.html}{interactive version} reveals the corresponding country name and metric value.}
  \label{fig:ungrouped-dissimilarity-plot}
\end{figure}

In the interactive version of Figure \ref{fig:ungrouped-dissimilarity-plot} available via \url{https://oluwayomi-olaitan.github.io/rjournal-wdiexplorer-interactive-plots/interactive-plots/ungrouped_dissimilarity_plot.html}, hovering over each highlighted line displays the country name and the average dissimilarity distance value. This plot visually complements and reinforces the earlier findings from the \texttt{pm\_variation} output generated by the \texttt{compute\_variation} function in Subsection 5.2. Qatar, Niger displayed in {\color{purple-blue} purple-blue} and Mauritania displayed in {\color{purple} reddish-purple} are the top three countries with the highest average dissimilarity distance. As seen in Figure \ref{fig:ungrouped-dissimilarity-plot}, these three countries also record the highest annual PM$_{2.5}$ air pollution exposure levels, with Qatar leading. Qatar's extensive oil and gas production and large-scale construction activities likely contribute to its elevated pollution levels and its distinctiveness relative to other countries. \cite{ali2018determination} identified Qatar arid climate, along with rapid industrialisation, urbanisation, and traffic emissions, as major contributing factors to its elevated PM$_{2.5}$ and PM$_{10}$ pollution levels.

\begin{verbatim}
# grouped version
plot_data_trajectories(
        pm_data, 
        metric_summary = pm_diagnostic_metrics_group, 
        metric_var = "within_group_avg_dist",
        group_var = "region"
)
\end{verbatim}

\begin{figure}[H]
  \centering
  \includegraphics[width=\textwidth]{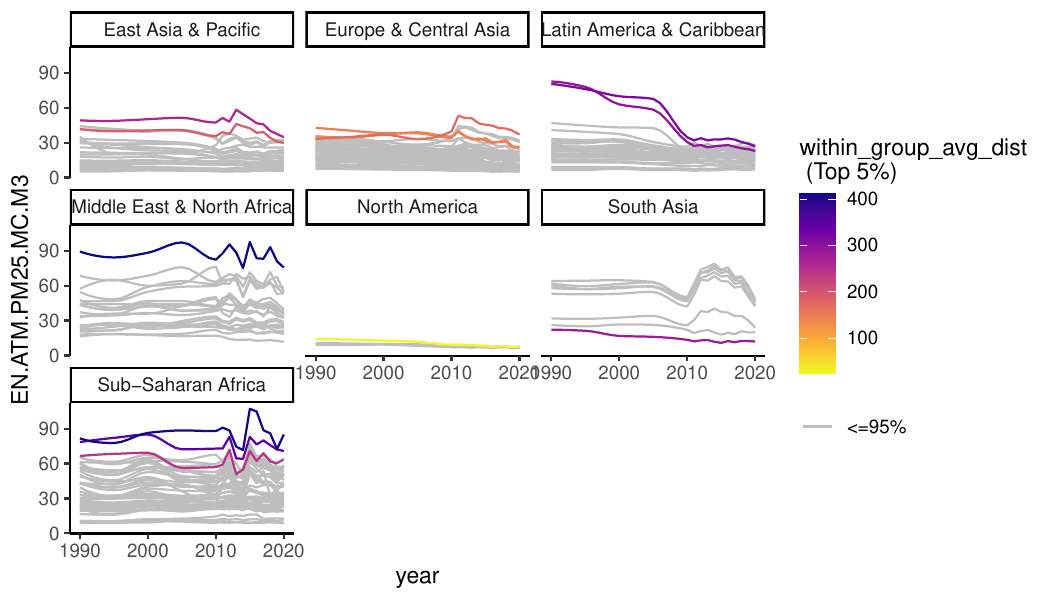}
  \caption{PM$_{2.5}$ air pollution data trajectories faceted by region groupings with group-based threshold rather than a uniform global threshold for highlighting countries with the top percentile. Qatar stood out with the highest dissimilarity values across other countries in Middle East \& North Africa while Niger, Mauritania and Senegal are identified as countries with the highest dissimilarity within the Sub-Saharan Africa region. Hovering any of the highlighted lines in the \href{https://oluwayomi-olaitan.github.io/rjournal-wdiexplorer-interactive-plots/interactive-plots/grouped_dissimilarity_plot.html}{interactive version} reveals the corresponding country name and metric value.}
  \label{fig:grouped-dissimilarity-plot}
\end{figure}

The interactive version of  Figure \ref{fig:grouped-dissimilarity-plot} is available online via \url{https://oluwayomi-olaitan.github.io/rjournal-wdiexplorer-interactive-plots/interactive-plots/grouped_dissimilarity_plot.html}. Figure \ref{fig:grouped-dissimilarity-plot} shows that within each region, countries are highlighted based on group-specific thresholds. Qatar stands out among all other countries in Middle East \& North Africa while Niger and Mauritania stand out in Sub-Saharan Africa. This suggests that their pollution trajectory over time are not only unusual at the global level but also relative to other countries in their region. In the study conducted by \cite{bauer2019desert}, air pollution across the African continent was found to be dominated by Saharan dust, with biomass burning identified as a major contributor, particularly in Central and West Africa. These natural environmental factors largely explain the high PM$_{2.5}$ air pollution levels and structural distinctiveness observed in Niger and Mauritania.

In Figure \ref{fig:ungrouped-dissimilarity-plot}, countries highlighted based on the global threshold include Afghanistan, Bahrain, Egypt, Arab Rep., India, Mali, Mauritania, Niger, Peru, Qatar, and Senegal. However, in Figure \ref{fig:grouped-dissimilarity-plot} where countries are highlighted based on the group-specific threshold, some of these countries are no longer highlighted, while new ones appear within their respective regions. Notably, no countries from East Asia \& Pacific or North America region were highlighted under the global threshold, yet region-specific highlighting reveals outliers that would otherwise be overlooked. This shows the value of group-specific thresholds in revealing regionally unusual temporal behaviours.

The results from the country average and within-group average dissimilarity align with the patterns observed in the silhouette width partition plot. Countries such as Qatar, Niger, and Mauritania were identified as outliers not only due to their high country average and within-group average dissimilarity, they also belong to regions that were characterised by weak similarity and within group variability. In contrast, Bermuda, Canada and the United States, all in the North America region, exhibit high positive silhouette widths, indicating relatively strong similarity among the countries in the group. However, the group-specific dissimilarity threshold highlights United States as a country with the highest dissimilarity within the region. Although the countries appear relatively similar, the data series for the United States differs slightly from the data series of Bermuda and Canada. This is also reflected in its silhouette width, which is positive but lower than the other countries in the North America region. The alignment across these measures underscores the value of using multiple metrics to identify structural patterns and detect outliers in country-level panel data.

Having introduced the chosen diagnostic indices, the code below demonstrate how the \\ \texttt{plot\_data\_trajectories} function can be used to highlight countries with the strongest linear trends (positive and negative), based on the absolute value of the linearity measure.

\begin{verbatim}
pm_trend_shape <- pm_trend_shape |>
      dplyr::mutate(abs_linearity = abs(linearity))

plot_data_trajectories(
        pm_data, 
        metric_summary = pm_trend_shape, 
        metric_var = "abs_linearity",
        percentile = 0.96
    )
\end{verbatim}

\begin{figure}[H]
  \centering
  \includegraphics[width=\textwidth]{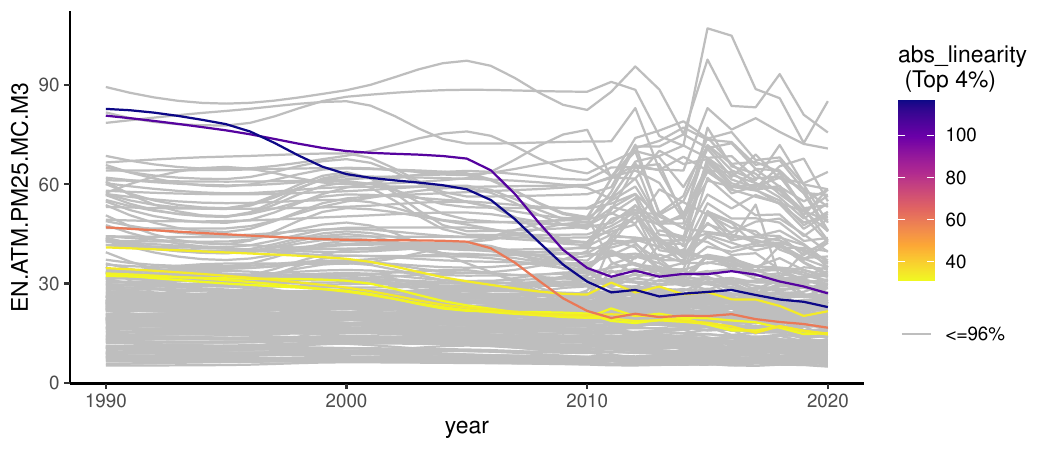}
  \caption{Ungrouped PM$_{2.5}$ air pollution data trajectories with highlighted countries based on the linearity metric values. Countries with absolute linearity values below or at the 96th percentile are displayed in grey, while countries within the top 4\% absolute linearity values are displayed using a colour gradient. All highlighted countries have negative linear trend, indicating that the most pronounced linear trends in the data reflect strong decline in the decomposed series. Hovering any of the highlighted lines in the \href{https://oluwayomi-olaitan.github.io/rjournal-wdiexplorer-interactive-plots/interactive-plots/ungrouped_linearity_plot.html}{interactive version} reveals the corresponding country name and metric value.}
  \label{fig:ungrouped-linearity-plot}
\end{figure}

The interactive version of Figure \ref{fig:ungrouped-linearity-plot} is available online at \url{https://oluwayomi-olaitan.github.io/rjournal-wdiexplorer-interactive-plots/interactive-plots/ungrouped_linearity_plot.html}. Bolivia, Peru were already identified as top 2 countries with highest negative linearity value in Subsection 5.2, along with other countries falling within the top 4\% are highlighted using a colour gradient. In Figure \ref{fig:ungrouped-linearity-plot}, the highlighted countries exhibit negative linear trend. This shows that, the most extreme linearity values ($\beta_1$) of the decomposed trend component have sustained downward trajectories. Figure \ref{fig:ungrouped-distribution-plot} supports this as the distribution plot of the linearity metric tends to be left skewed with majority of the countries exhibiting negative linearity values. \cite{sicard2023trends} identify that the dominating decrease in air pollution is primarily attributed to significant reductions in emissions and the establishment and implementation of environmental policies and legislation. According to \cite{xu2023factors}, The Gross Domestic Product (GDP) per capita has influenced the shift of PM$_{2.5}$ air pollution from positive to negative in most countries by $2010$, reflecting increased investment in environmental protection and the introduction of air pollution prevention and control bills. Figure \ref{fig:ungrouped-linearity-plot} confirms these reductions in most countries.

\begin{verbatim}
pm_trend_shape_group <- add_group_info(
      metric_summary = pm_trend_shape, 
      pm_data
  )
  
plot_data_trajectories(
        pm_data, 
        metric_summary = pm_trend_shape_group, 
        metric_var = "abs_linearity",
        group_var = "region",
        percentile = 0.96
    )
\end{verbatim}

\begin{figure}[H]
  \centering
  \includegraphics[width=\textwidth]{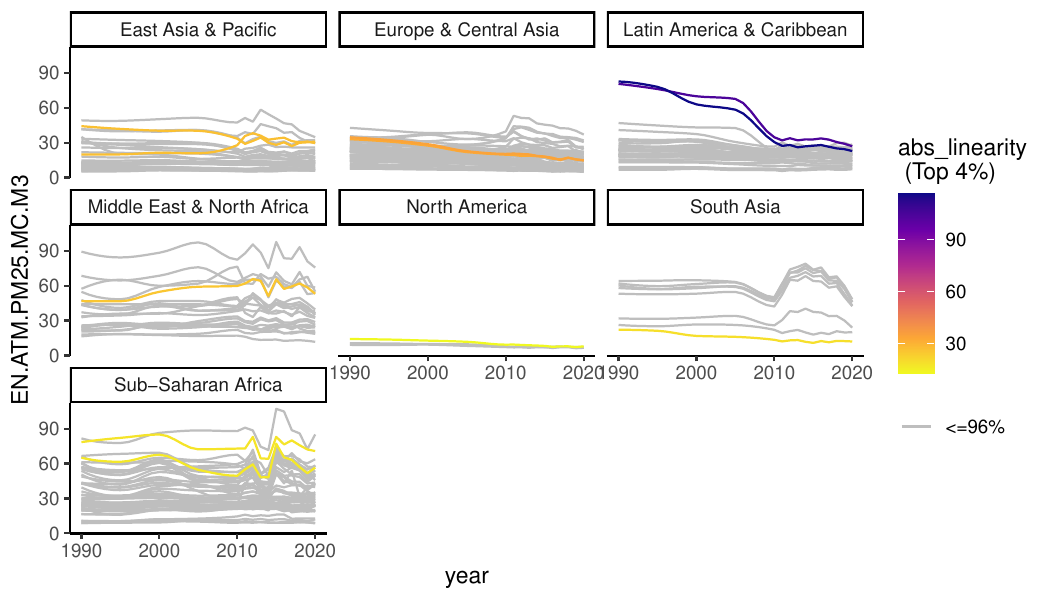}
  \caption{PM$_{2.5}$ air pollution data trajectories faceted by region groupings with group-based threshold with highlighted countries based on their linearity metric values. Countries with absolute linearity values below or at the 96th percentile are displayed in grey, while countries within the top 4\% absolute linearity values are displayed using a colour gradient. Hovering any of the highlighted lines in the \href{https://oluwayomi-olaitan.github.io/rjournal-wdiexplorer-interactive-plots/interactive-plots/grouped_linearity_plot.html}{interactive version} reveals the corresponding country name and metric value.}
  \label{fig:grouped-linearity-plot}
\end{figure}

In Figure \ref{fig:grouped-linearity-plot}, countries are highlighted based on a group-specific thresholds, which reveal those exhibiting pronounced linear trends within each region. The interactive version of Figure \ref{fig:grouped-linearity-plot} available via \url{https://oluwayomi-olaitan.github.io/rjournal-wdiexplorer-interactive-plots/interactive-plots/grouped_linearity_plot.html} shows that in the East Asia \& Pacific region, Mongolia and Thailand are the highlighted countries. Mongolia demonstrates a positive linear trend while Thailand exhibits a negative trend. This indicates that, though the ungrouped version shows that highlighted countries within the top 4\% absolute linearity values all demonstrates negative trends, within the East Asia \& Pacific region shows otherwise. Countries falling within the 96th percentile exhibit positive and negative trends. In Europe \& Central Asia, both Moldova and Ukraine show clear decreasing linear trends in their PM$_{2.5}$ data series. Peru and Bolivia in Latin America \& Caribbean, have the most pronounced decreasing trend. The United States in North America; Maldives in South Asia; Mauritania and Nigeria in Sub-Saharan Africa also exhibit decreasing trends. In contrast, Saudi Arabia in Middle East \& North Africa displays an increasing linear trend.

These observed decreasing trends in PM$_{2.5}$ air pollution levels align with the findings of \cite{guerreiro2014air}, who reported substantial reductions in particulate matter emissions with significant improvements between $2000$ and $2019$ in regions such as Eastern United States, Europe, Southeast China, and Japan. These reductions are linked to strong regulatory frameworks and air quality initiatives implemented in these areas, contributing to declining particulate matter concentrations. Precipitation also plays a role through wet deposition, acting as an efficient mechanism for pollutant removal \cite{xu2023factors}. However, while significant progress is noted in developed areas, challenges remain in developing regions, such as India and the Middle East, where PM$_{2.5}$ concentrations often continue to rise due to the prioritisation of economic development over environmental concerns.

\subsubsection{Parallel Coordinates Plot}

The \texttt{plot\_parallel\_coords} function generates parallel coordinate plots with normalised metric values to a scale of $0$ to $1$. It takes two main arguments, a data frame \texttt{diagnostic\_summary} containing all diagnostic indices values alongside the pre-defined grouping information, and a variable specified by \texttt{colour\_var} in the data frame used to assign colours to the parallel lines. If an additional optional argument \texttt{group\_var} is specified, the function instead produces a grouped parallel coordinate plot, where metric values are normalised within each group before plotting, and the resulting plot is faceted by the specified grouping variable. The parallel coordinate plot is rendered interactively using \textit{ggiraph} with tooltips to display the corresponding country name, metric and metric value of each parallel line. 

\begin{verbatim}
# ungrouped version
plot_parallel_coords(
      diagnostic_summary = pm_diagnostic_metrics_group,
      colour_var = "region"
)
\end{verbatim}

\begin{figure}[H]
  \centering
  \includegraphics[width=\textwidth]{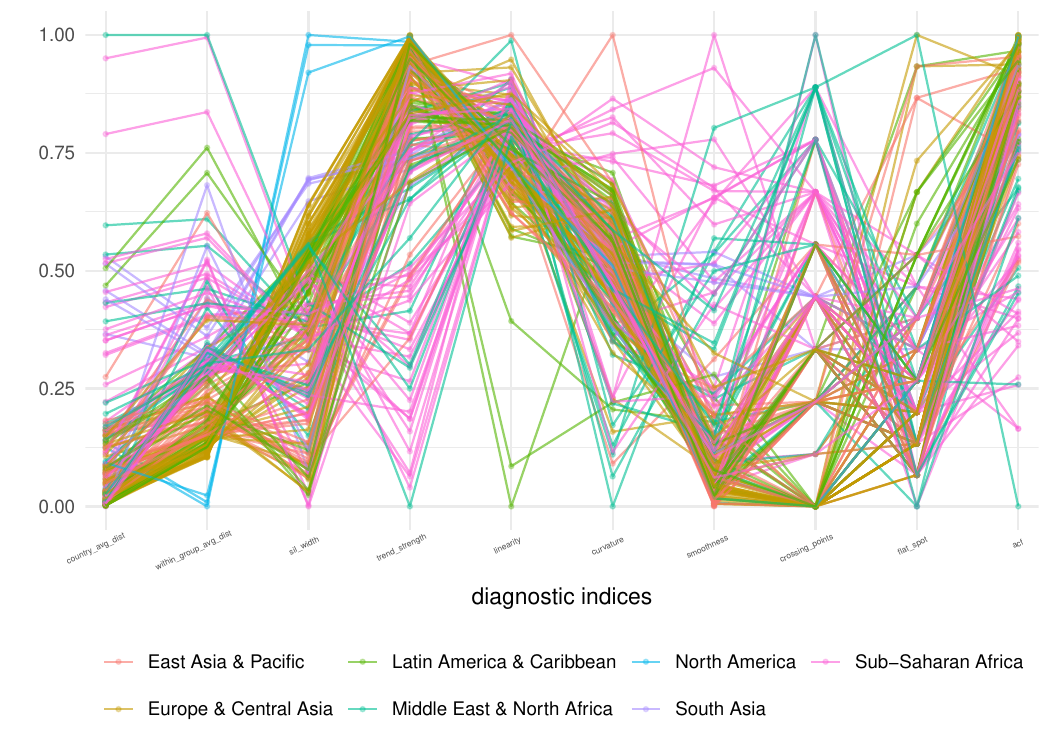}
  \caption{Parallel coordinate plot displaying the metric values across all the diagnostic indices. The metric values are normalised to a scale of $0$ to $1$. Countries in Sub-Saharan Africa region, shown in {\color{magenta}magenta}, display a wide spread across most diagnostics indices. Hovering any parallel line in the \href{https://oluwayomi-olaitan.github.io/rjournal-wdiexplorer-interactive-plots/interactive-plots/ungrouped-parallel-plot.html}{interactive version} reveals the corresponding country name and hovering the points across the x-axis reveals the metric name, metric value and the country name.}
  \label{fig:ungrouped-parallel-plot}
\end{figure}

The interactive version of Figure \ref{fig:ungrouped-parallel-plot} available via \url{https://oluwayomi-olaitan.github.io/rjournal-wdiexplorer-interactive-plots/interactive-plots/ungrouped_parallel_plot.html} displays the parallel coordinates across all 10 diagnostic indices. Hovering over the x-axis, the tooltips show the country name of each parallel line, the correspondence metric, and its metric value. This collectively reinforce the findings presented in Subsection 5.2.

\begin{verbatim}
# grouped version
plot_parallel_coords(
      diagnostic_summary = pm_diagnostic_metrics_group,
      colour_var = "region",
      group_var = "region"
)
\end{verbatim}

\begin{figure}[H]
  \centering
  \includegraphics[width=\textwidth]{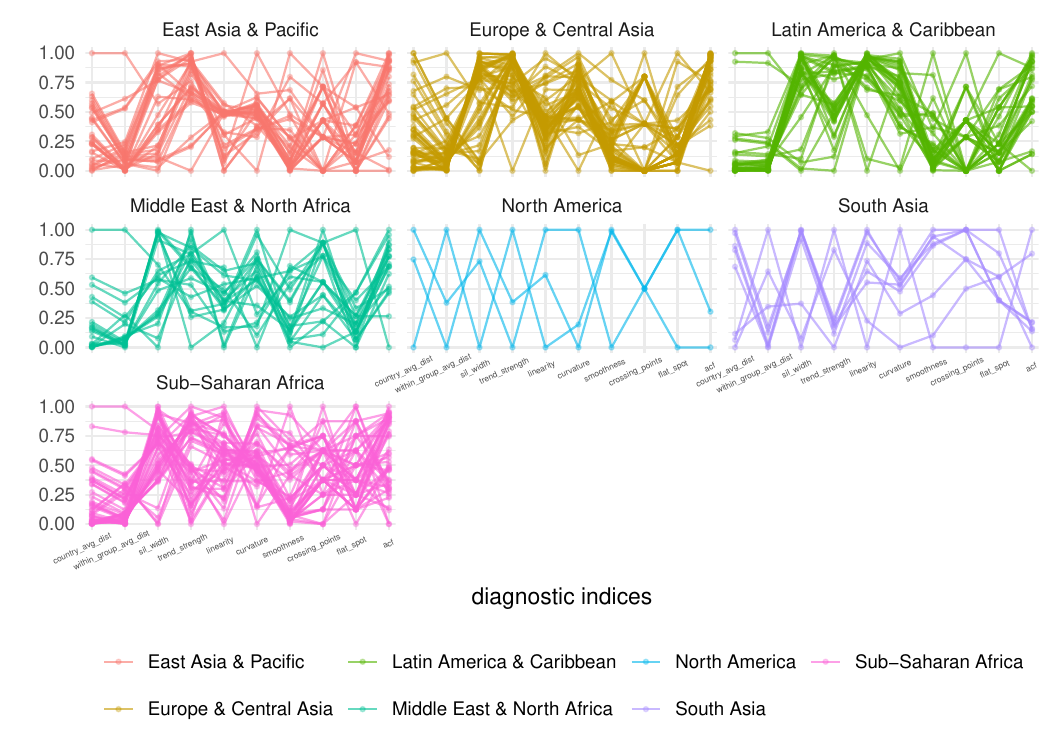}
  \caption{Parallel coordinate plot displaying the metric values across all diagnostic indices grouped by region. The metric values are normalised to a scale of $0$ to $1$ within each group. Countries in Sub-Saharan Africa region, shown in {\color{magenta}magenta}, displays the widest spread across most diagnostics indices. Hovering any parallel line in the \href{https://oluwayomi-olaitan.github.io/rjournal-wdiexplorer-interactive-plots/interactive-plots/grouped-parallel-plot.html}{interactive version} reveals the corresponding country name and hovering the points across the x-axis reveals the metric name, metric value and the country name.}
  \label{fig:grouped-parallel-plot}
\end{figure}

Figure \ref{fig:ungrouped-parallel-plot} reveals that countries in North America, shown in {\color{skyblue}skyblue} consistently recorded almost similar values, with slight difference in United States reinforcing earlier findings about the behavioural patterns of PM$_{2.5}$ air pollution levels within this region, see \ref{fig:grouped-parallel-plot} with its interactive version available via \url{https://oluwayomi-olaitan.github.io/rjournal-wdiexplorer-interactive-plots/interactive-plots/grouped_parallel_plot.html}. In North America region their silhouette widths are all positive and high as seen in Figure \ref{fig:partition-plot}, further supporting these findings. These countries exhibit high trend strength, coupled with negative linearity and close to $0$ curvature, indicating a clear, consistent, and dominant linear decline in PM$_{2.5}$ levels over time, with the United States showing the most pronounced downward trend. Their low smoothness scores further imply that the time series evolve gradually without abrupt changes. The data series of each of these countries crosses its median only once, pointing to a single shift in their series followed by sustained stability. Both Canada and the United States exhibit 7 long flat spots, while Bermuda exhibits 5, indicating periods where their data series remained within an interval. This is consistent with the overall smoothness and linearity of their trends. Additionally, they all show high autocorrelation, with the United States having the highest value, reflecting strong temporal correlation where each series is closely related to its preceding and succeeding time points. 

The countries in Sub-Saharan Africa region, shown in {\color{magenta}magenta}, display a widespread index values. In contrast, countries in Europe \& Central Asia region, represented in {\color{darkgoldenrod2}darkgoldenrod}, exhibit relatively uniform behaviour in most of the diagnostic indices.

United Arab Emirates in the Middle East \& North Africa region, shown in {\color{tealgreen}tealgreen} is another country that stands out in Figure \ref{fig:grouped-parallel-plot}, primarily due to its distinctive values across several diagnostic metrics. Notably, the data series records low values for both trend strength and autocorrelation, indicating that it exhibit minimal to no consistent temporal pattern and successive observations are largely independent of one another. However, it simultaneously records some of the highest values for the number of crossing points and the longest flat spot metrics. These suggest extended periods of stability with shifts around its median values. 

Bolivia, a country in the Latin America \& Caribbean region, shown in {\color{leafgreen}leafgreen}  also emerges as a notable country that stands out in both  Figures \ref{fig:ungrouped-parallel-plot} and \ref{fig:grouped-parallel-plot}, characterised by a unique combination of high trend strength, low linearity, low crossing points, and high autocorrelation. The high trend strength indicates a pronounced directional movement in its PM$_{2.5}$ air pollution data series, while the low linearity value confirms that this movement is a consistent linear decline. The low number of crossing points (to be precise $1$) implies that the data series crosses its median value once, possibly due to a sustained downward linear trend with no subsequent directional changes. The high autocorrelation reflects strong correlation between each series and its preceding time point. This shows that, the PM$_{2.5}$ air pollution levels over time in Bolivia follow a stable and strongly consistent downward linear trend.

\subsubsection{Link-view of Metrics and Series Plot}

The \texttt{plot\_metric\_linkview} function creates an interactive linked dual-based visualisation that connects the diagnostic space with the corresponding data trajectories. It takes three main arguments: a dataset containing data for a selected WDI indicator, \texttt{metric\_summary}, a data frame containing the computed diagnostic indices with the pre-defined grouping information ; and \texttt{metric\_var} a pair of metric variables within the \texttt{metric\_summary} data frame which are used to create a scatterplot. The function also accepts an optional \texttt{index} argument which defaults to $\mathrm{NULL}$ and the function uses the first variable stored as an attribute in the WDI dataset. By default, the function generates an ungrouped interactive link-based visualisation. However, if an additional optional argument \texttt{group\_var} is specified, the function instead produces a grouped version of the interactive dual-based visualisation of a scatterplot of two selected diagnostic metrics alongside their series trajectories, faceted by the specified grouping variable.

\begin{verbatim}
# ungrouped version
plot_metric_linkview(
          pm_data, 
          metric_summary = pm_diagnostic_metrics,
          metric_var = c("linearity", "curvature")
        )
\end{verbatim}

\begin{figure}[H]
  \centering
  \includegraphics[width=\textwidth]{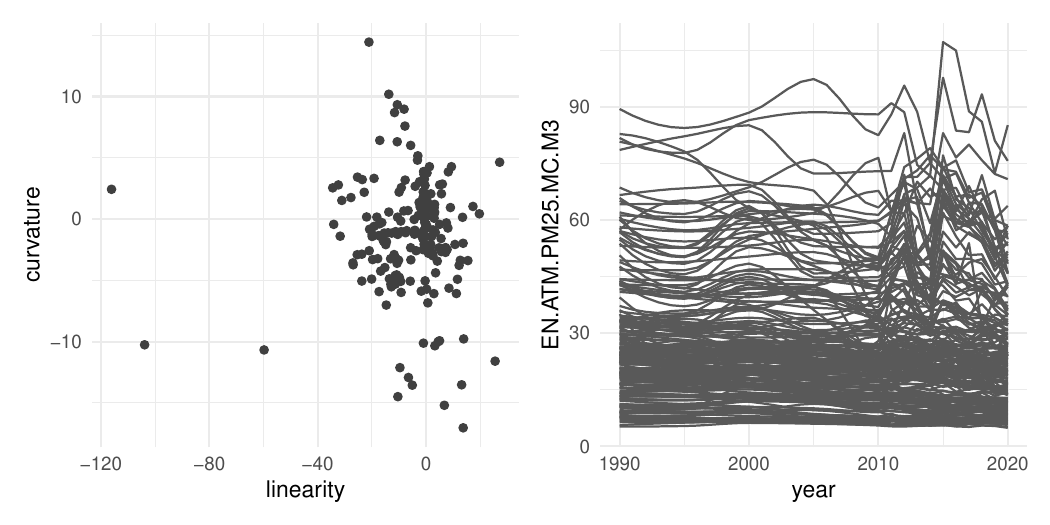}
  \caption{Link-based plot showing the relationship between linearity and curvature metrics across all countries. Each point in the scatterplot represents a country, and hovering a point in the \href{https://oluwayomi-olaitan.github.io/rjournal-wdiexplorer-interactive-plots/interactive-plots/ungrouped_linkview_plot.html}{interactive version} reveals the corresponding trajectory of the data series.}
  \label{fig:ungrouped-linkview-plot}
\end{figure}

The interactive version of Figure \ref{fig:ungrouped-linkview-plot} available via \url{https://oluwayomi-olaitan.github.io/rjournal-wdiexplorer-interactive-plots/interactive-plots/ungrouped_linkview_plot.html} simultaneously displays a link view of the metric values and their correspondence data trajectories: one panel displays data series line plots for all countries, while the other shows a scatterplot of the relationship between the linearity and curvature metrics. This enables an interactive exploration of viewing the metrics values in the data space. Hovering over a line in the time series plot highlights the corresponding country and its scatter point value in the other plot, and vice versa. This dynamic interaction enhances an intuitive understanding of how temporal behaviours relate to pair of structural features of the data, captured by the collection of diagnostic indices.

\begin{verbatim}
# grouped version
plot_metric_linkview(
          pm_data, 
          metric_summary = pm_diagnostic_metrics_group,
          metric_var = c("acf", "flat_spot"),
          group_var = "region"
      ) 
\end{verbatim}

\begin{figure}[H]
  \centering
  \includegraphics[width=\textwidth]{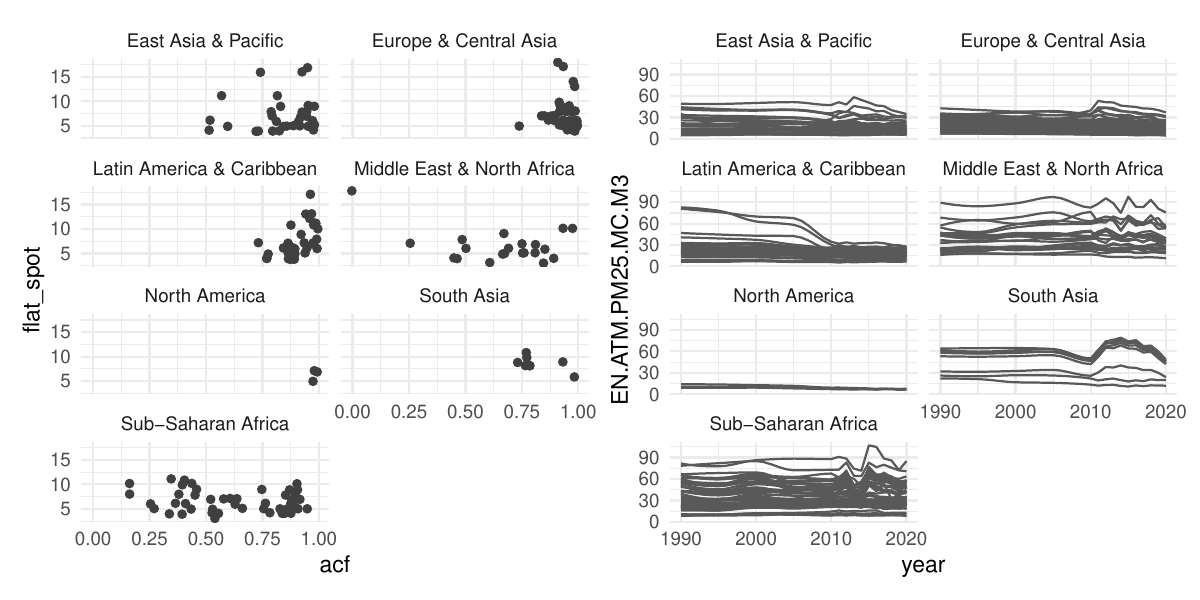}
  \caption{Link-based plot showing the relationship between autocorrelation and the longest flat spot metrics across all countries faceted by region. Each point in the scatterplot represents a country, and hovering a point in the \href{https://oluwayomi-olaitan.github.io/rjournal-wdiexplorer-interactive-plots/interactive-plots/grouped_linkview_plot.html}{interactive version} reveals the corresponding trajectory of the data series in its panel.}
  \label{fig:grouped-linkview-plot}
\end{figure}

Figure \ref{fig:grouped-linkview-plot} with its interactive version available via \url{https://oluwayomi-olaitan.github.io/rjournal-wdiexplorer-interactive-plots/interactive-plots/grouped_linkview_plot.html} extends this view by incorporating group-level structure through faceting to the scatterplot of autocorrelation and the longest flat spot metrics and data trajectories plot. In this version, the scatterplot and data trajectories panels are faceted by region, allowing within-group comparisons of metrics relationships and temporal behaviours. This facilitates the identification of countries that behave differently from others within the same group, based on both their diagnostic indices and data trajectories.

The link-view plot is highly flexible, allowing users to visualise the scatterplot of any two selected diagnostic metrics and the data trajectories plot. Extending this approach, a linked version of the parallel coordinates plot connected to the trajectory plot would further enhance interpretability by enabling simultaneous exploration of multiple metrics for each country and how they reflect in the temporal series.

\section{Discussion}\label{discussion}
The primary contribution of \textit{wdiexplorer} lies in enabling enhanced exploratory analysis of country-level panel data within the R programming environment. Its functionality specifically supports the need to analyse the temporal behaviour of WDI data, both at the level of individual countries and across countries of the same peer group. Furthermore, the incorporation of data grouping structures in visually presenting the analytical results facilitates the identification of unique patterns, outliers, and other interesting characteristics in temporal behaviours. This integration reinforces a comprehensive workflow where numerical assessment and visual representation enhance better understanding and interpretation of the data.

\textit{wdiexplorer}, as the acronym implies ``World Development Indicators explorer'', is primarily designed for the exploration of WDI country-level panel data, but it is not limited to this database. The package can also accommodate other databases with data characterised by repeated observations of the same variables collected over time from multiple countries passed directly into the package functions, provided the \texttt{index} variable is specified across all functions that requires the variable with the country-year numerical values. In additional, the computational functions of \textit{wdiexplorer} that compute the diagnostic metrics can be integrated with a wide range of visualisation tools. This flexibility enables users to explore the metrics values within custom visualisations, while enabling visual exploration of the outputs to the underlying quantitative diagnostic measures.

Several functions have been implemented in the \textit{wdiexplorer} package to support (1) data sourcing and preparation, (2) computation of the set of diagnostic indices (3) both static and interactive visualisations. The workflow organises the package functions into these three stages to provide a seamless and user-friendly process. As illustrated in Section \ref{workflow}, we demonstrated how to source and prepare the data, use the available 14 functions and interpret the results at every stage of the workflow. To ensure comparability, all functions were applied to the PM$_{2.5}$ air pollution indicator data and interpreted accordingly. These features establish \textit{wdiexplorer} as a self-contained R package for the exploration of country-level panel data.

We provide users with practical guidance to facilitate reproducibility through comprehensive vignettes available via \url{https://oluwayomi-olaitan.github.io/wdiexplorer/articles/pm_analysis.html}. This vignette demonstrates how to apply each of the 14 functions of the \textit{wdiexplorer} package to explore World Development Indicators data. It guides users through example workflows for two indicators: one with a complete annual dataset (PM$_{2.5}$ air pollution data) and another with missing entries and data collected triennially (PISA data). It contains interpretations of the diagnostic indices output and demonstrates the visualisation tools, highlighting their analytical and visualisation capabilities.

A potential direction for future work is to extend the static visualisation functions to support interactivity. The current implementation of \texttt{plot\_metric\_distribution} generates a static representation of the distribution of metric values, summarising the variation by displaying the spread and shape of their distributions, with individual countries represented by dots. Enhancing this function with interactive capabilities will enable users to hover over a point to reveal the corresponding country name in the distribution plot. This would align the \texttt{plot\_metric\_distribution} function with other interactive visual functions available in the \textit{wdiexplorer} package.

In conclusion, the \textit{wdiexplorer} package contribute to advancing a comprehensive workflow for exploratory analysis of country-level panel data. By following our step-by-step workflow, researchers and other R users can better understand their data, identify patterns and outliers, and communicate results in the context of the data structure more effectively.

\section{Acknowledgments}\label{acknowledgments}

This publication has emanated from research conducted with the financial support of Taighde Éireann – Research Ireland under Grant number 18/CRT/6049. For the purpose of Open Access, the author has applied a CC BY public copyright licence to any Author Accepted Manuscript version arising from this submission.


\newpage

\bibliographystyle{unsrt}  
\bibliography{wdiexplorer}

@Manual{R,
    title = {R: A Language and Environment for Statistical Computing},
    author = {{R Core Team}},
    organization = {R Foundation for Statistical Computing},
    address = {Vienna, Austria},
    year = {2024},
    url = {https://www.R-project.org/},
  }

@dataset{worldbank-wdi,
  author       = {{The World Bank}},
  title        = {{World Development Indicators (WDI)}},
  year         = {2025},
  publisher    = {World Bank},
  howpublished = {\url{https://databank.worldbank.org/source/world-development-indicators}},
  note         = {Accessed: 2025-08-06}
}

@misc{worldbank_classifier,
  author       = {{The World Bank}},
  title        = {How does the World Bank classify countries?},
  year         = 2025,
  url          = {https://datahelpdesk.worldbank.org/knowledgebase/articles/378834-how-does-the-world-bank-classify-countries},
  note         = {Accessed: 2025-08-06}
}

@misc{worldbank_country_lending,
  author       = {{The World Bank}},
  title        = {World Bank Country and Lending Groups},
  year         = {2025},
  url          = {https://datahelpdesk.worldbank.org/knowledgebase/articles/906519-world-bank-country-and},
  note         = {Accessed: 2025-08-08}
}

@book{gelman2007data,
  title={Data analysis using regression and multilevel/hierarchical models},
  author={Gelman, Andrew and Hill, Jennifer},
  year={2007},
  publisher={Cambridge university press}
}

@article{RJ-2021-050,
  author = {Wang, Earo and Cook, Dianne},
  title = {Conversations in Time: Interactive Visualization to Explore Structured Temporal Data},
  journal = {The R Journal},
  year = {2021},
  note = {https://doi.org/10.32614/RJ-2021-050},
  doi = {10.32614/RJ-2021-050},
  volume = {13},
  issue = {1},
  issn = {2073-4859},
  pages = {461-469}
}

@article{qu2024exploratory,
  title={Exploratory functional data analysis},
  author={Qu, Zhuo and Dai, Wenlin and Euan, Carolina and Sun, Ying and Genton, Marc G},
  journal={Test},
  pages={1--24},
  year={2024},
  publisher={Springer}
}

@inproceedings{dang2014scagexplorer,
  title={Scagexplorer: Exploring scatterplots by their scagnostics},
  author={Dang, Tuan Nhon and Wilkinson, Leland},
  booktitle={2014 IEEE Pacific visualization symposium},
  pages={73--80},
  year={2014},
  organization={IEEE}
}

@inproceedings{tukey1985computer,
  title={Computer graphics and exploratory data analysis: An introduction},
  author={Tukey, John W and Tukey, Paul A},
  booktitle={Proceedings of the sixth annual conference and exposition: computer graphics},
  volume={85},
  number={3},
  pages={773--785},
  year={1985}
}

@article{wilkinson2008scagnostics,
  title={Scagnostics distributions},
  author={Wilkinson, Leland and Wills, Graham},
  journal={Journal of Computational and Graphical Statistics},
  volume={17},
  number={2},
  pages={473--491},
  year={2008},
  publisher={Taylor \& Francis}
}

@article{rousseeuw1987silhouettes,
  title={Silhouettes: a graphical aid to the interpretation and validation of cluster analysis},
  author={Rousseeuw, Peter J},
  journal={Journal of computational and applied mathematics},
  volume={20},
  pages={53--65},
  year={1987},
  publisher={Elsevier}
}

@article{luedicke2015friedman,
  title={Friedman’s super smoother},
  author={Luedicke, Joerg},
  journal={Boston College. http://fmwww. bc. edu/RePEc/bocode/s/supsmooth\_doc. pdf},
  year={2015}
}

@book{hyndman2021forecasting,
  author    = {Hyndman, Rob J. and Athanasopoulos, George},
  title     = {Forecasting: Principles and Practice},
  publisher = {OTexts},
  year      = {2021},
  edition   = {3rd},
  address   = {Melbourne, Australia},
  url       = {https://OTexts.com/fpp3},
  note      = {Accessed on 2025-07-05}
}

@inproceedings{ali2018determination,
  title={Determination of the levels of particulate matter 25 and 10 and their elemental composition in Qatar},
  author={Ali Ahmadi, Ahmad and Balakrishnan, Perumal and Kakosimos, Konstantinos and Goktepe, Ipek},
  booktitle={Qatar Foundation Annual Research Conference Proceedings},
  volume={2018},
  pages={EEPD912},
  year={2018},
  organization={HBKU Press Qatar}
}

@article{bauer2019desert,
  title={Desert dust, industrialization, and agricultural fires: Health impacts of outdoor air pollution in Africa},
  author={Bauer, Susanne E and Im, Ulas and Mezuman, Keren and Gao, Chloe Y},
  journal={Journal of Geophysical Research: Atmospheres},
  volume={124},
  number={7},
  pages={4104--4120},
  year={2019},
  publisher={Wiley Online Library}
}

@article{sicard2023trends,
  title={Trends in urban air pollution over the last two decades: A global perspective},
  author={Sicard, Pierre and Agathokleous, Evgenios and Anenberg, Susan C and De Marco, Alessandra and Paoletti, Elena and Calatayud, Vicent},
  journal={Science of The Total Environment},
  volume={858},
  pages={160064},
  year={2023},
  publisher={Elsevier}
}

@article{xu2023factors,
  title={What factors dominate the change of PM2. 5 in the world from 2000 to 2019? A study from multi-source data},
  author={Xu, Xiankang and Shi, Kaifang and Huang, Zhongyu and Shen, Jingwei},
  journal={International journal of environmental research and public health},
  volume={20},
  number={3},
  pages={2282},
  year={2023},
  publisher={MDPI}
}

@article{guerreiro2014air,
  title={Air quality status and trends in Europe},
  author={Guerreiro, Cristina BB and Foltescu, Valentin and De Leeuw, Frank},
  journal={Atmospheric environment},
  volume={98},
  pages={376--384},
  year={2014},
  publisher={Elsevier}
}

@article{cleveland1990stl,
  title={STL: A seasonal-trend decomposition},
  author={Cleveland, Robert B and Cleveland, William S and McRae, Jean E and Terpenning, Irma and others},
  journal={J. off. Stat},
  volume={6},
  number={1},
  pages={3--73},
  year={1990}
}

@Manual{WDI,
    title = {WDI: World Development Indicators and Other World Bank Data},
    author = {Vincent Arel-Bundock},
    year = {2025},
    note = {R package version 2.7.9},
    url = {https://CRAN.R-project.org/package=WDI},
    doi = {10.32614/CRAN.package.WDI},
  }

@Manual{wbstats,
    title = {wbstats: Programmatic Access to the World Bank API},
    author = {Jesse Piburn},
    organization = {Oak Ridge National Laboratory},
    address = {Oak Ridge, Tennessee},
    year = {2020},
    url = {https://doi.org/10.11578/dc.20171025.1827},
  }

@Manual{worldbank,
    title = {worldbank: Client for World Banks's 'Indicators' and 'Poverty and
Inequality Platform (PIP)' APIs},
    author = {Maximilian Mücke},
    year = {2025},
    note = {R package version 0.6.0},
    url = {https://CRAN.R-project.org/package=worldbank},
    doi = {10.32614/CRAN.package.worldbank},
  }

@Article{brolgar,
    title = {The R Journal: brolgar: An R package to BRowse Over Longitudinal Data Graphically and Analytically in R},
    author = {Nicholas Tierney and Di Cook and Tania Prvan},
    journal = {The R Journal},
    year = {2022},
    volume = {14},
    issue = {2},
    pages = {6-25},
    note = {https://doi.org/10.32614/RJ-2022-023},
    issn = {2073-4859},
    doi = {10.32614/RJ-2022-023},
  }

@Manual{tsibbletalk,
    title = {tsibbletalk: Interactive Graphics for Tsibble Objects},
    author = {Earo Wang and Di Cook},
    year = {2020},
    note = {R package version 0.1.0},
    url = {https://CRAN.R-project.org/package=tsibbletalk},
    doi = {10.32614/CRAN.package.tsibbletalk},
  }

@Article{tsibble,
    author = {Earo Wang and Dianne Cook and Rob J Hyndman},
    title = {A new tidy data structure to support exploration and modeling of temporal data},
    journal = {Journal of Computational and Graphical Statistics},
    volume = {29},
    number = {3},
    pages = {466-478},
    year = {2020},
    publisher = {Taylor & Francis},
    doi = {10.1080/10618600.2019.1695624},
    url = {https://doi.org/10.1080/10618600.2019.1695624},
  }

@Manual{crosstalk,
    title = {crosstalk: Inter-Widget Interactivity for HTML Widgets},
    author = {Joe Cheng and Carson Sievert},
    year = {2023},
    note = {R package version 1.2.1},
    url = {https://CRAN.R-project.org/package=crosstalk},
  }

@Manual{feasts,
    title = {feasts: Feature Extraction and Statistics for Time Series},
    author = {Mitchell O'Hara-Wild and Rob Hyndman and Earo Wang},
    year = {2024},
    note = {R package version 0.4.1},
    url = {https://CRAN.R-project.org/package=feasts},
  }

@Manual{fabletools,
    title = {fabletools: Core Tools for Packages in the 'fable' Framework},
    author = {Mitchell O'Hara-Wild and Rob Hyndman and Earo Wang},
    year = {2024},
    note = {R package version 0.5.0},
    url = {https://CRAN.R-project.org/package=fabletools},
    doi = {10.32614/CRAN.package.fabletools},
  }

@Book{ggplot2,
    author = {Hadley Wickham},
    title = {ggplot2: Elegant Graphics for Data Analysis},
    publisher = {Springer-Verlag New York},
    year = {2016},
    isbn = {978-3-319-24277-4},
    url = {https://ggplot2.tidyverse.org},
  }

@Manual{ggiraph,
    title = {ggiraph: Make 'ggplot2' Graphics Interactive},
    author = {David Gohel and Panagiotis Skintzos},
    year = {2025},
    note = {R package version 0.8.12},
    url = {https://CRAN.R-project.org/package=ggiraph},
  }

@Manual{ggdist,
    title = {{ggdist}: Visualizations of Distributions and Uncertainty},
    author = {Matthew Kay},
    year = {2024},
    note = {R package version 3.3.2},
    url = {https://mjskay.github.io/ggdist/},
    doi = {10.5281/zenodo.3879620},
  }

@Article{naniar,
    title = {Expanding Tidy Data Principles to Facilitate Missing Data Exploration, Visualization and Assessment of Imputations},
    author = {Nicholas Tierney and Dianne Cook},
    journal = {Journal of Statistical Software},
    year = {2023},
    volume = {105},
    number = {7},
    pages = {1--31},
    doi = {10.18637/jss.v105.i07},
  }

@Manual{scales,
    title = {scales: Scale Functions for Visualization},
    author = {Hadley Wickham and Thomas Lin Pedersen and Dana Seidel},
    year = {2025},
    note = {R package version 1.4.0},
    url = {https://CRAN.R-project.org/package=scales},
  }

@Book{flexclust,
    author = {Sara Dolnicar and Bettina Grün and Friedrich Leisch},
    title = {Market Segmentation Analysis---Understanding It, Doing It, and Making It Useful},
    publisher = {Springer},
    year = {2018},
    address = {Singapore},
    doi = {10.1007/978-981-10-8818-6},
  }

\end{document}